\documentclass[11pt,journal,onecolumn,draftcls]{IEEEtran}
\usepackage [dvips]{graphicx}
\usepackage{subfig}
\usepackage{psfrag}
\usepackage{multirow}

\newcommand{\bd}{\ensuremath{\mathbf{d}}}
\newcommand{\mD}{\ensuremath{\mathcal{D}}}

\newcommand{\mC}{\ensuremath{\mathcal{C}}}

\newtheorem{theorem}{Theorem}
\newtheorem{lemma}{Lemma}
\newtheorem{corollary}{Corollary}
\newtheorem{definition}{Definition}
\newtheorem{proposition}{Proposition}
\newtheorem{remark}{Remark}

\usepackage[cmex10]{amsmath}
\usepackage{amssymb}

\begin{document}

\title{Completion Time in Two-user Channels: An Information-Theoretic Perspective}
\author{Yuanpeng~Liu,
        Elza~Erkip,~\IEEEmembership{Fellow,~IEEE}%
\thanks{The material in this paper was presented in part at the 49th Annual Allerton Conference on Communication, Control, and Computing and 2011 IEEE Information Theory Workshop.}%
\thanks{Y. Liu was with the Department of Electrical and Computer Engineering, Polytechnic School of Engineering, New York University, Brooklyn, NY, USA (e-mail: yl755@nyu.edu).}%
\thanks{E. Erkip is with the Department of Electrical and Computer Engineering, Polytechnic School of Engineering, New York University, Brooklyn, NY, USA (e-mail: elza@nyu.edu).}}

\maketitle

\begin{abstract}
In a two-user channel, completion time refers to the number of channel uses spent by each user to transmit a bit pool with some given size. In this paper, the information-theoretic formulation of completion time is based on the concept of constrained rates, where users are allowed to employ different numbers of channel uses for transmission as opposed to the equal channel use of the standard information-theoretic formulation. Analogous to the capacity region, the completion time region characterizes all possible trade-offs among users' completion times. For a multi-access channel, it is shown that the completion time region is achieved by operating the channel in two independent phases: a multi-access phase when both users are transmitting, and a point-to-point phase when one user has finished and the other is still transmitting. Using a similar two-phase approach, the completion time region (or inner and outer bounds) is established for a Gaussian broadcast channel and a Gaussian interference channel. It is observed that although consisting of two convex subregions, the completion time region may not be convex in general. Finally an optimization problem of minimizing the weighted sum completion time for a Gaussian multi-access channel and a Gaussian broadcast channel is solved, demonstrating the utility of the completion time approach.
\end{abstract}

\section{Introduction}
The information-theoretic study of a communication network design is usually guided by the assumption that users' data buffers are always full. This assumption, on one hand, greatly simplifies the problem and enables a rigorous mathematical methodology to study networks. On the other hand, it ignores the bursty nature of real sources and lacks delay considerations, leading to the so-called unconsummated union between information theory and communication networks \cite{Ephremides}. There has been a substantial body of literature that made progress in bridging the gap between information theory and communication networks, for example see \cite{Telatar}-\cite{Minero}. However, a large portion of the problem remains unresolved. In particular, delay, as argued in \cite{Ephremides}, is not only a performance metric, but more importantly a fundamental quantity that may also affect the communication rate-accuracy trade-off.

The most natural way to model transmission delay is to simply view it as a function of rate, specifically the total number of bits divided by rate. Suppose one is interested in studying the minimum sum delay in a two-user Gaussian symmetric multi-access channel with each user transmitting a fairly large amount of data, say $\tau$ bits. Following the above approach, one would formulate an optimization problem
\begin{align*}
    \textrm{minimize    }&\quad \frac{\tau}{r_1}+\frac{\tau}{r_2}\\
    \textrm{subject to  }&\quad (r_1,r_2)\in\mathcal{C},
\end{align*}
where $\mathcal{C}=\left\{(r_1,r_2):  0\leq \sum_{i\in \Omega}r_i\leq \gamma(\sum_{i\in \Omega}P_i),\ \Omega\subseteq \{1,2\}\right\}$, $\gamma(P)=\frac{1}{2}\log(1+P)$, is the capacity region of the multi-access channel. Accordingly the solution is to let each user transmit at half of the sum rate defining $\mathcal{C}$, yielding a total sum delay of $4\tau/\gamma(2P)$. However there exists an alternative scheme that outperforms the above strategy. First the two users could transmit at the corner point of $\mathcal{C}$ until one user has finished transmitting of its bits. Then the remaining user boosts its rate to the point-to-point capacity to complete transmission of its remaining bits. This results in a total sum delay of $4\tau/\gamma(P)-\tau\gamma(2P)/\gamma^2(P)$, which is less than $4\tau/\gamma(2P)$. This example illustrates that the limitation of modeling delay as a function of the multi-access capacity region stems from the very assumption underlying the formulation of this region, i.e. users' data buffers are always full and hence there is always multi-user interference.

In this paper, we study a specific communication scenario --- transmitting large files over two-user channels. Our goal is to study transmission delay in an information-theoretic setting through investigating the \textit{completion time} problem. Mathematically we model the large file transmission as follows: for each user, there are $m_i\tau_i$, $i=1,2$, bits to be transmitted where $m_i$ is a scaling factor to ensure information-theoretic arguments with large block lengths can be invoked. Let $n_i$ be the actual number of channel uses for user $i$ to complete the transmission. Then the \textit{normalized completion time} (or \textit{completion time} for short) is defined as $n_i/m_i$ in the limit of large $n_i$ and $m_i$.

Assuming no transmitter/receiver cooperation or feedback, we focus on three representative classes of two-user channels: multi-access channel (MAC), broadcast channel (BC) and interference channel (IC). We formulate completion time using the concept of \textit{constrained rates}, where users' codewords are constrained to span different block-lengths. This allows us to relax the full-buffer assumption and hence leads to a way of incorporating delay in an information-theoretic setting. The achievability of completion time is defined in terms of the achievability of the corresponding constrained rate. In order to obtain the completion time region $\mathcal{D}^*$, i.e. the set of all achievable completion times, we show that it is necessary to obtain the corresponding constrained capacity region. We first consider a discrete memoryless multi-access channel (DM-MAC) and explicitly characterize its constrained capacity region. In the achievable scheme, one operates the channel in two independent phases: a multi-access phase when both users are transmitting subject to the standard DM-MAC rate constraint and a point-to-point phase when one user has finished and the other transmits at the maximum point-to-point rate. For a Gaussian multi-access channel (GMAC) assuming expected per symbol power constraints, we establish a closed-form expression of $\mathcal{D}^*$. For a broadcast channel and an interference channel, since even the standard capacity region remains unknown for a generic discrete memoryless case \cite{El Gamal}, we only focus on the Gaussian channels. We derive a closed-form expression of the completion time region of a Gaussian broadcast channel (GBC) and find inner and outer bounds of the completion time region for a Gaussian interference channel (GIC). Throughout these investigations, we observe that although consisting of two convex subregions, the completion time region as a whole may not be convex. To demonstrate the usage of the completion time region, we seek one particular utility optimization---the weighted sum completion time minimization over $\mathcal{D}^*$---for a GMAC and a GBC. We also extend the discussion for Gaussian channels to the case where the expected per symbol power constraint is replaced by the expected block power constraint.

One key observation made in this paper is that operating the channel in two independent phases --- multi-user phase and single-user phase --- depending on transmission completion status suffices to achieve the completion time region. This decoupling greatly simplifies the codebook design as in each phase, the standard capacity achieving codebook can be used and there is no need to code across phases. While the completion time formulation can be readily extended to more complicated channels such as channels with relays, transmitter/receiver cooperations and feedbacks, we provide a simple example of a Gaussian Z interference channel with a relay to demonstrate that the multi-phase scheme may be no longer optimal. We leave the analysis of completion time in more complicated multi-user channels with cooperation and/or feedback to future studies.

This work, in many aspects, is inspired by \cite{Rai}, where the authors solved the sum completion time minimization problem for a $K$-user symmetric GMAC by drawing an analogy to multi-processor queues. Compared to \cite{Rai}, we give a general formulation of the completion time and provide a complete characterization of the completion time region for a two-user DM-MAC, GMAC and GBC. In another related work \cite{Ng}, the authors considered an interference channel where power control is used to minimize some convex utility function over the completion time region obtained by treating interference as noise. Compared to \cite{Ng}, we adopt an information-theoretic approach without restriction to any specific coding scheme such as treating interference as noise.

This paper is organized as the follows. In Section II, the concept of constrained rates is introduced, based on which a formulation of completion time is then given. Section III treats a DM-MAC and a GMAC, where a detailed derivation of the completion time region is given. Following the same approach, the completion time regions of a GBC and a GIC are obtained in Section IV. Section V discusses utility optimization using the completion time region and Section VI provides an extension for Gaussian channels with the expected block power constraint. Conclusions and discussions are provided in Section VII.

\emph{Notation}: The logarithm is with respect to base 2. We let $\gamma(x)=\frac{1}{2}\log(1+x)$, $[X]^+=\max\{X,0\}$. We denote the empty set by $\Phi$. In comparing two vectors,  $\mathbf{x}=(x_1,...,x_n)$, $\mathbf{y}=(y_1,...,y_n)$, $\mathbf{x}\leq \mathbf{y}$ means $x_i\leq y_i$ for all $i\in\{1,...,n\}$.

\section{Problem Formulation And Preliminaries}
\label{section:formulation}
We assume the total amount of information to be transmitted for user $i$ is $m_i\tau_i$ bits with $\tau_i$ fixed, $i=1,2$. Here $m_i$ is an asymptotically large scaling factor used to ensure information theoretic arguments. For example, this could correspond to each user having a large file to communicate. Unlike the classical information theory, which is based on a full-buffer assumption, having some users finish early is not only desirable to reduce the completion time of those users, but also preferable for the remaining users since they can enjoy reduced multi-user interference for the remaining transmission. In order to capture this, and to formulate the completion time problem, we will define communication rates over different number of channel uses for different users, as opposed to the standard definition, where users' codewords span the same number of channel uses. We refer this as {\em constrained rate} and provide a complete definition for MAC in Section II.A. Similar definitions for BC and IC are briefly discussed in Section II.B and II.C, where the differences are highlighted. We only discuss discrete memoryless models, definitions for their Gaussian counterparts can be given similarly. In Section II.D, completion time is formally defined and its relation to constrained rate is established.

In the following, we let $n_i$ be the number of actual channel uses for user $i$, $i\in\{1,2\}$. We denote $n=\max\{n_1,n_2\}$ where we let $n_i\rightarrow\infty$ with $c=\lim_{n_i\rightarrow\infty} n_1/n_2$ so that information-theoretic arguments with large block lengths can be invoked. The analysis of completion time depends on the order of user transmission completion. Hence we define:
\begin{align}
\label{def:index}
  \pi_1=\arg_{i=1,2}\min\{n_i\},\ \pi_2=\arg_{i=1,2}\max\{n_i\}.
\end{align}

\subsection{Constrained Rate Region for Multi-Access Channel}
Consider a discrete memoryless multi-access channel (DM-MAC) $(\mathcal{X}_1\times\mathcal{X}_2,p(y|x_1,x_2),\mathcal{Y})$, where $\mathcal{X}_1,\mathcal{X}_2$ are the input alphabets, $\mathcal{Y}$ is the output alphabet and $p(y|x_1,x_2)$ is the channel transition probability. Let $X_{i,t}\in\mathcal{X}$ be the channel input for user $i$ at time $t$ and $Y_t\in\mathcal{Y}$ be the channel output at time $t$. Associated with each user is a message set $\mathcal{W}_i=\{1,...,M_i\}$, from which message $W_i\in\mathcal{W}$ is randomly drawn with a uniform distribution. Let $\mathbf{n}=(n_1,n_2)$ and $\mathbf{M}=(M_1,M_2)$.
\begin{definition}
\label{def:code}
An $(\mathbf{M},\mathbf{n})$ \textit{constrained code} consists of the following:
\begin{enumerate}
\item Message sets: $\mathcal{W}_i=\{1,...,M_i\},\quad i=1,2$
\item A set of encoding functions:
\begin{align*}
    X_{\pi_1,t}:\begin{cases}
		  \mathcal{W}_{\pi_1} \rightarrow \mathcal{X}_{\pi_1},\quad 0\leq t\leq n_{\pi_1}\\
		  X_{\pi_1,t}=\psi_{\pi_1,t},\quad n_{\pi_1}< t \leq n
                \end{cases},\qquad X_{\pi_2,t}: \mathcal{W}_{\pi_2} \rightarrow \mathcal{X}_{\pi_2}
\end{align*}
Note that user $\pi_1$ transmits $\psi_{\pi_1,t}$ after it completes information transmission. The sequence $\Psi_{\pi_1}=\{\psi_{\pi_1,t}:\psi_{\pi_1,t}\in\mathcal{X}_{\pi_1}, t>n_{\pi_1}\}$ is announced to all nodes prior to communication and carries no information.
\item A set of decoding functions: $\widehat{W}_i:\mathcal{Y}^{n_i}\rightarrow \mathcal{W}_i,\quad i=1,2$
\end{enumerate}
\end{definition}

The error probability of an $(\mathbf{M},\mathbf{n})$ constrained code is then given by
\begin{align*}
    P_e(\mathbf{n})=P_r\left( \widehat{W}_1\neq W_1\textrm{ or } \widehat{W}_2\neq W_2 \right).
\end{align*}
\begin{definition}
\label{def:rate}
For a family of $(\mathbf{M},\mathbf{n})$ constrained codes with fixed $c=\lim_{n_i\rightarrow\infty} n_1/n_2$, the \textit{$c$-constrained rates} $\mathbf{R}=(R_1,R_2)$ are defined as
\begin{equation*}
    R_i=\lim_{n_i\rightarrow\infty}\frac{1}{n_i}\log M_i, \quad i=1,2.
\end{equation*}
\end{definition}

\begin{definition}
\label{achievableratedef}
Pair $(\mathbf{R},c)$ is said to be \textit{achievable} if there exists a sequence of $(\mathbf{M},\mathbf{n})$ constrained codes such that $P_e(\mathbf{n})\rightarrow 0$ as $n_i\rightarrow \infty$, $i=1,2,$ with $c=\lim_{n_i\rightarrow\infty} n_1/n_2$. For a given $c$, we define the \textit{c-constrained rate region}, denoted by $\mathfrak{R}_{c}$, as the set of $\mathbf{R}$ such that $(\mathbf{R},c)$ is achievable for some given coding scheme. Similarly for a given $c$, the \textit{c-constrained capacity region}, denoted by $\mathfrak{C}_{c}$, is the closure of the set of achievable $\mathbf{R}$.
\end{definition}

\begin{remark}
\label{remark:constrained}
We use the term ``c-constrained rate/capacity region'' to emphasize the fact that the effective ratio of the codeword lengths is $c$ and rates $(R_1,R_2)$ are defined accordingly. Consequently $\mathfrak{R}_{c}$ and $\mathfrak{C}_{c}$ are functions of $c$. Using this denotation, $\mathfrak{R}_{1}$ and $\mathfrak{C}_{1}$ correspond to the standard rate region and capacity region respectively. To keep the notation concise, for the rest of this paper, the term ``rate/capacity region'' refers to the standard rate/capacity region, which is denoted by $\mathcal{R}$ and $\mathcal{C}$ respectively, i.e. $\mathcal{R}=\mathfrak{R}_{1}$ and $\mathcal{C}=\mathfrak{C}_{1}$.
\end{remark}

\subsection{Constrained Rate Region for Broadcast Channel}
Consider a discrete memoryless broadcast channel (DM-BC) $(\mathcal{X},p(y_1,y_2|x),\mathcal{Y}_1\times\mathcal{Y}_2)$ with individual message sets $\mathcal{W}_i=\{1,...,M_i\}$, $i=1,2$, where $\mathcal{X}$ is the input alphabet, $\mathcal{Y}_1,\mathcal{Y}_2$ are the output alphabets and $p(y_1,y_2|x)$ is the channel transition probability. Message $W_i$ is randomly drawn from $\mathcal{W}_i$ with a uniform distribution.
\begin{definition}
An $(\mathbf{M},\mathbf{n})$ constrained code consists of the following:
\begin{enumerate}
\item Message sets: $\mathcal{W}_i=\{1,...,M_i\},\quad i=1,2$
\item An encoding function:
\begin{align*}
    X_{t}:\begin{cases}
		  (\mathcal{W}_1\times\mathcal{W}_2) \rightarrow \mathcal{X},\quad 0\leq t\leq n_{\pi_1}\\
		  \mathcal{W}_{\pi_2} \rightarrow \mathcal{X},\quad n_{\pi_1}< t \leq n
                \end{cases}
\end{align*}
Note that the channel input is determined jointly by both users' messages for $t\leq n_{\pi_1}$, but it is determined solely by user $\pi_2$'s message for $n_{\pi_1}< t \leq n$.
\item A set of decoding functions: $\widehat{W}_i:\mathcal{Y}_i^{n_i}\rightarrow \mathcal{W}_i,\quad i=1,2$
\end{enumerate}
\end{definition}
The remaining definitions follow those in Section II.A.

\subsection{Constrained Rate for Interference Channel}
Consider a discrete memoryless interference channel (DM-IC) $(\mathcal{X}_1\times\mathcal{X}_2,p(y_1,y_2|x_1,x_2), \mathcal{Y}_1\times \mathcal{Y}_2)$, where $\mathcal{X}_1,\mathcal{X}_2$ are the input alphabets, $\mathcal{Y}_1,\mathcal{Y}_2$ are the output alphabets and $p(y_1,y_2|x_1,x_2)$ is the channel transition probability. Associated with each user is a message set $\mathcal{W}_i=\{1,...,M_i\}$, from which message $W_i\in\mathcal{W}$ is randomly drawn with a uniform distribution.
\begin{definition}
An $(\mathbf{M},\mathbf{n})$ constrained code consists of the following:
\begin{enumerate}
\item Message sets: $\mathcal{W}_i=\{1,...,M_i\},\quad i=1,2$
\item A set of encoding functions:
\begin{align*}
    X_{\pi_1,t}:\begin{cases}
		  \mathcal{W}_{\pi_1} \rightarrow \mathcal{X}_{\pi_1},\quad 0\leq t\leq n_{\pi_1}\\
		  X_{\pi_1,t}=\psi_{\pi_1,t},\quad n_{\pi_1}< t \leq n
                \end{cases},\qquad X_{\pi_2,t}: \mathcal{W}_{\pi_2} \rightarrow \mathcal{X}_{\pi_2}
\end{align*}
Note that user $\pi_1$ transmits $\psi_{\pi_1,t}$ at time $t$ after it completes information transmission. The sequence $\Psi_{\pi_1}=\{\psi_{\pi_1,t}:\psi_{\pi_1,t}\in\mathcal{X}_{\pi_1}, t>n_{\pi_1}\}$ is announced to all nodes prior to communication and carries no information.
\item A set of decoding functions: $\widehat{W}_i:\mathcal{Y}_i^{n_i}\rightarrow \mathcal{W}_i,\quad i=1,2$
\end{enumerate}
\end{definition}
The above definition is very similar to Definition \ref{def:code} for a multi-access channel with the only difference being in the decoding functions. The remaining definitions follow those in Section II.A.

\subsection{Completion Time}
\begin{definition}
Consider a two-user channel that is either DM-MAC, DM-BC or DM-IC. Suppose we have a sequence of $(\mathbf{M},\mathbf{n})$ constrained codes with $\log M_i=m_i\tau_i$, $i=1,2$, where $\tau_i$ is fixed and $m_i=\Theta(n_i)$. Then \textit{normalized completion time} ( or \textit{completion time} for short ) for user $i$ is defined as $d_i=\lim_{n_i\rightarrow\infty}n_i/m_i$.
\end{definition}

\begin{definition}
\label{originaldef}
For a given $\tau=(\tau_1,\tau_2)$, completion time $\mathbf{d}_{\tau}=(d_1,d_2)$ is said to be \textit{achievable} if there exists a sequence of $(\mathbf{M},\mathbf{n})$ constrained codes with completion times $(d_1,d_2)$ such that $P_e(\mathbf{n})\rightarrow 0$ as $n_i\rightarrow \infty$, $i=1,2$. The \textit{achievable completion time region}, denoted by $\mathcal{D}_{\tau}$, is the set of achievable $\mathbf{d}_{\tau}$ for a given coding scheme. Similarly the \textit{completion time region}, denoted by $\mathcal{D}^*_{\tau}$, is the closure of the set of achievable completion times $\mathbf{d}_{\tau}$.
\end{definition}

For conciseness, in the rest of the paper we drop the subscript $\tau$ and simply refer completion time as $\mathbf{d}$ and (achievable) completion time region as $\mathcal{D}^*$ ($\mathcal{D}$), but it is clear that these quantities are functions of $\tau$. The following lemma implies that the achievability of completion time is equivalent to the achievability of constrained rates defined earlier.

\begin{lemma}
\label{ctiff}
Completion time $\mathbf{d}$ is achievable iff $(\mathbf{R}, c)$ is achievable, where
\begin{align}
\label{ctdefc}
\mathbf{R}=\left(\frac{\tau_1}{d_1},\frac{\tau_2}{d_2}\right),\quad c=\frac{d_1}{d_2}.
\end{align}
\end{lemma}

\begin{IEEEproof}
For the if part, recall Definition \ref{achievableratedef}. The pair $(\mathbf{R},c)$ is achievable if there exists a sequence of $(\mathbf{M},\mathbf{n})$ constrained codes such that $P_e(\mathbf{n})\rightarrow 0$ as $n_i\rightarrow \infty$ with $c=\lim_{n_i\rightarrow\infty}n_1/n_2$. With $\mathbf{R}$ in (\ref{ctdefc}) and Definition \ref{def:rate}, we have, for large $n_i$, $\log M_i=n_iR_i =\frac{n_i}{d_i}\tau_i=m_i\tau_i$, where we let $m_i\triangleq\frac{n_i}{d_i}$. Therefore by Definition \ref{originaldef}, the sequence of $(\mathbf{M},\mathbf{n})$ constrained codes that achieves $(\mathbf{R},c)$ with $\mathbf{R}$, $c$ given by (\ref{ctdefc}) also achieves $\mathbf{d}$.

For the only if part, we consider Definition \ref{originaldef}: Completion time $\mathbf{d}$ is achievable if there exists a sequence of $(\mathbf{M},\mathbf{n})$ constrained codes such that $\log M_i=m_i\tau_i$ and $P_e(\mathbf{n})\rightarrow 0$ with $d_i=\lim_{n_i\rightarrow\infty}n_i/m_i$. Without loss of generality, we can set $m=m_1=m_2$. Then we have $\lim_{n_i\rightarrow\infty}\frac{n_1}{n_2}=\frac{d_1}{d_2}=c$ and $\lim_{n_i\rightarrow\infty}\frac{1}{n_i}\log M_i=\frac{\tau_i}{d_i}$, $i=1,2$. Therefore by Definition \ref{achievableratedef} the sequence of $(\mathbf{M},\mathbf{n})$ constrained codes that achieves $\mathbf{d}$ also achieves $(\mathbf{R},c)$ with $\mathbf{R}$, $c$ given by (\ref{ctdefc}).
\end{IEEEproof}

The following corollary is a direct consequence of Definition \ref{originaldef} and Lemma \ref{ctiff}.
\begin{corollary}
\label{newdef}
$\mathcal{D}=\left\{\mathbf{d}: \mathbf{R}=(\frac{\tau_1}{d_1},\frac{\tau_2}{d_2}),c=\frac{d_1}{d_2},\mathbf{R}\in \mathfrak{R}_c\right\}$ is an achievable completion time region and $\mathcal{D}^*=\left\{\mathbf{d}: \mathbf{R}=(\frac{\tau_1}{d_1},\frac{\tau_2}{d_2}),c=\frac{d_1}{d_2},\mathbf{R}\in \mathfrak{C}_{c}\right\}$ is the completion time region.
\end{corollary}

Corollary \ref{newdef} as it stands is not very useful for characterizing $\mathcal{D}^*$. An achievable $\mathbf{d}$ can be expressed in terms of achievable $c$-constrained rates $\mathbf{R}\in\mathfrak{C}_{c}$, where $\mathfrak{C}_{c}$ in return depends on $\mathbf{d}$ through $c$. This means that in order to obtain $\mathcal{D}^*$ --- the closure of the set of achievable $\mathbf{d}$ --- we have to consider not just one $\mathfrak{C}_{c}$ for a given $c$, but a family of regions which are in return parameterized by $\mathbf{d}$ in the set. Because of this dependence, it is easy to check whether or not a given $\mathbf{d}$ is achievable, provided $\mathfrak{C}_{c}$ as a function of $c$ is known, but difficult to compute $\mathcal{D}^*$ using Corollary \ref{newdef}. To circumvent this issue, in the Section III and Section IV we will use $\mathfrak{C}_{c}$ as a bridge to relate the completion time and the standard rate, from which we will eventually establish $\mathcal{D}^*$ for a GMAC and a GBC and inner/outer bounds of $\mathcal{D}^*$ for a GIC.

\section{Completion Time Region of a Multi-Access Channel}
In this section, we focus on a two-user multi-access channel. As discussed in Section II-D, it is difficult to directly determine $\mathcal{D}^*$ using Corollary \ref{newdef}. In this section, we adopt an indirect approach. Specifically, we will first obtain the $c$-constrained capacity region of a DM-MAC, based on which a map between the completion time region and the standard capacity region can be derived. We then argue that the completion time region consists of two convex subregions. Each subregion will be determined individually, where the convexity and the map are used together to argue the achievability and the converse. Finally, the union of the two subregions gives rise to the completion time region of a DM-MAC. To substantiate the discussion, we also consider a GMAC, where explicit characterization of $\mathcal{D}^*$ is obtained. These steps are applicable to broadcast channel and interference channel as well, which will be treated in Section IV.

\subsection{Constrained Capacity Region}
\begin{lemma}
\label{lem:acheivecrate}
For a DM-MAC, the $c$-constrained rates $\mathbf{R}=(R_1,R_2)$ are achievable, if
\begin{enumerate}
\item
for $c\leq 1$, $R_2$ can be decomposed into $R_2'$ and $R_2''$: $R_2=cR_2'+(1-c)R_2''$, such that $(R_1,R_2')\in\mathcal{C}$, $R_2''\leq r_2^*$,

\item
for $c\geq 1$, $R_1$ can be decomposed into $R_1'$ and $R_1''$: $R_1=\frac{1}{c}R_1'+(1-\frac{1}{c})R_1''$, such that $(R_1',R_2)\in\mathcal{C}$, $R_1''\leq r_1^*$,
\end{enumerate}
where $\mathcal{C}$ is the capacity region of a DM-MAC given by
\begin{align}
\label{eq:maccapa}
    \mathcal{C} = \left\{ \begin{aligned} &(r_1,r_2): (r_1,r_2)\in\mathbb{R}_2^+, r_1\leq I(X_1,Y|X_2,Q), r_2\leq I(X_2,Y|X_1,Q)\\
    &r_1+r_2\leq I(X_1,X_2;Y|Q), \textrm{ for some } p(q)p(x_1|q)p(x_2|q)p(y|x_1,x_2)
 \end{aligned}\right\},
\end{align}
and $r_i^*$ is the point-to-point capacity given by, for $i,j\in\{1,2\},i\neq j$,
\begin{align}
    r_i^*&=\max_{p_{X_i}} I(X_i;Y|X_j=\psi_j^*), \label{def:ratemac} \\
    \psi_j^*&= \arg\max_{\psi\in\mathcal{X}_j} \max_{p_{X_i}} I(X_i;Y|X_j=\psi). \label{eq:psi}
\end{align}
\end{lemma}

\begin{IEEEproof}
We prove for $c\leq 1$. The case of $c\geq 1$ follows similarly. Consider a time sharing scheme: for the first $n_1$ channel uses, a multi-user codebook achieving $(R_1,R_2')\in\mathcal{C}$ is used; for the remaining $n_2-n_1$ channel uses, a single-user codebook achieving $R_2''\leq r_{2}^*$ is used for user 2 while user 1 transmits a constant symbol $\psi_1^*$. Since in each sub-interval error probability can be made arbitrarily small, the overall time sharing scheme results in vanishing probability of error. User 2's overall rate is then given by $R_2=\log M_2/n_2=[n_1R_2'+(n_2-n_1)R_2'']/n_2=cR_2'+(1-c)R_2''$. Therefore $(R_1,R_2)$ is an achievable $c$-constrained rate pair.
\end{IEEEproof}

\begin{remark}
It is important to realize that the constrained rate for each user is only defined over the channel uses during which the user is active. In the above example, user 1's rate is defined over the first $n_1$ channel uses, after which it transmits a constant symbol $\psi_1^*$ which ``opens'' up the channel the most to facilitate user 2's remaining transmission. Also note that we choose lower case $r$ to indicate standard rate and upper case $R$ for constrained rate to distinguish the two quantities. This convention will be used in the remainder of this paper.
\end{remark}

\begin{theorem}
\label{theo:network}
The $c$-constrained capacity region $\mathfrak{C}_c$ of a DM-MAC is the set of all non-negative $(R_1,R_2)$ satisfying
\begin{enumerate}
\item
$(R_1,[\frac{1}{c}R_2-(\frac{1}{c}-1)r_2^*]^+)\in\mathcal{C}$ for $c\leq 1$,
\item
$([cR_1-(c-1)r_1^*]^+,R_2)\in\mathcal{C}$ for $c\geq 1$,
\end{enumerate}
where $\mathcal{C}$ and $r_i^*$, $i\in\{1,2\}$, are given in (\ref{eq:maccapa}) and (\ref{def:ratemac}) respectively.
\end{theorem}

\begin{IEEEproof}	
We prove the theorem for $c\leq 1$. The case of $c\geq 1$ follows similarly. The achievability follows from Lemma \ref{lem:acheivecrate}. For $R_2\geq(1-c)r_2^*$, let $R_2''= r_2^*$ and $(R_1,R_2)$ is achievable if $(R_1,\frac{1}{c}R_2-(\frac{1}{c}-1)r_2^*)\in\mathcal{C}$. For $R_2<(1-c)r_2^*$, setting $R_2'=0$, $R_2''=\frac{1}{1-c}R_2$ (hence $R_2''<r_2^*$), we have $(R_1,R_2)$ is achievable if $(R_1,0)\in\mathcal{C}$. To conclude, $(R_1,R_2)$ is achievable if $(R_1,[\frac{1}{c}R_2-(\frac{1}{c}-1)r_2^*]^+)\in\mathcal{C}$.

For the converse, it is easy to see that the following inequalities constitute a multi-letter upper bound:
\begin{align*}
  n_1R_1-\epsilon_n &\leq I(W_1;Y^{n_1}),\\
  n_2R_2-\epsilon_n &\leq I(W_2;Y^{n_1}) + I(X_{2,n_1+1}^{n_2};Y_{n_1+1}^{n_2}|X_{1,n_1+1}^{n_2} = \psi_{1,n_1+1}^{n_2} ), \\
  n_1R_1+n_2R_2-\epsilon_n &\leq I(W_1,W_2;Y^{n_1}) + I(X_{2,n_1+1}^{n_2};Y_{n_1+1}^{n_2}|X_{1,n_1+1}^{n_2}= \psi_{1,n_1+1}^{n_2}),
\end{align*}
for some sequence $\Psi_1=\{\psi_{1,t}:\psi_{1,t} \in\mathcal{X}_1, t\in\{n_1+1,...,n_2\}\}$, where $\lim_{n_i\rightarrow \infty}\epsilon_n/n_i=0$.

The first terms of the RHS of these inequalities correspond to the standard multi-letter upper bounds for a DM-MAC, which can be further single-letterized following standard steps \cite[Chapter 15.3.4]{Cover}. Also, it is easy to argue $I(X_{2,n_1+1}^{n_2};Y_{n_1+1}^{n_2}|X_{1,n_1+1}^{n_2} = \psi_{1,n_1+1}^{n_2} ) \leq (n_2-n_1) r_2^*$, where the single-letterization for a point to point channel is used. Overall we obtain
\begin{align*}
  n_1R_1-\epsilon_n &\leq  n_1I(X_1;Y|X_2,Q)\\
  n_2R_2-\epsilon_n &\leq  n_1I(X_2;Y|X_1,Q) + (n_2-n_1)r_2^*\\
  n_1R_1+n_2R_2-\epsilon_n &\leq n_1I(X_1,X_2;Y|Q) + (n_2-n_1)r_2^*,
\end{align*}
where $Q$ is a uniformly distributed random variable on $\{1,...,n_1\}$. After dividing both sides with $n_1$ and moving $r_2^*$ terms to the left, we have
\begin{align*}
  R_1 &\leq  I(X_1;Y|X_2,Q)\\
  [\tfrac{1}{c}R_2-(\tfrac{1}{c}-1)r_2^*] &\leq  I(X_2;Y|X_1,Q)\\
  R_1+[\tfrac{1}{c}R_2- (\tfrac{1}{c}-1)r_2^*] &\leq I(X_1,X_2;Y|Q).
\end{align*}
This coincides with the achievable $c$-constrained rate region. Hence the $c$-constrained capacity region of a DM-MAC is established for $c\leq 1$.
\end{IEEEproof}

\subsection{Obtaining the Completion Time Region from the Capacity Region}
For a DM-MAC, we define two subregions of the completion time region $\mathcal{D}^*$ as
\begin{align}
\label{def:ctrsubregion}
  \mathcal{D}_1^*&=\mathcal{D}^*\bigcap\{(d_1,d_2):d_1\leq d_2\}, \quad \mathcal{D}_2^*=\mathcal{D}^*\bigcap\{(d_1,d_2):d_1\geq d_2\},
\end{align}
and two subregions of the capacity region $\mathcal{C}$ given in (\ref{eq:maccapa}) as
\begin{align}
\label{def:capasubregion}
\mathcal{C}_1=\mathcal{C}\bigcap \left\{(r_1,r_2): \tfrac{r_2}{r_1}\leq \tfrac{\tau_2}{\tau_1} \right\}, \quad \mathcal{C}_2=\mathcal{C}\bigcap \left\{(r_1,r_2): \tfrac{r_2}{r_1}\geq \tfrac{\tau_2}{\tau_1} \right\}.
\end{align}
Note that $\mathcal{C}_1$ and $\mathcal{C}_2$ are functions of $\tau_1$ and $\tau_2$.

\begin{lemma}
\label{lem:mapping}
For $\tau_i\neq 0$, consider a map $G_i:\mathbf{r}\rightarrow\mathbf{d}$ given by
\begin{align}
\label{lem:mapping:eqd1}
  d_i=\frac{\tau_i}{r_i},\quad d_j=\frac{\tau_j}{r_j^*}+\frac{(r_j^*-r_j)\tau_i}{r_j^*r_i},
\end{align}
for $i,j\in\{1,2\}$, $i\neq j$ and $r_j^*$ given in (\ref{def:ratemac}). Then $\mathbf{d}\in\mathcal{D}_i^*$ iff $\mathbf{d}=G_i(\mathbf{r})$ for some $\mathbf{r}\in\mathcal{C}_i$, $i\in\{1,2\}$.
\end{lemma}

\begin{IEEEproof}
In the following, we prove $G_1$. The case of $G_2$ follows similarly. We first show that $\mathbf{d}=G_1(\mathbf{r})$ for some $\mathbf{r}\in\mathcal{C}_1$ is achievable. Due to Lemma \ref{ctiff}, it suffices to show $\mathbf{R}=(\frac{\tau_1}{d_1}, \frac{\tau_2} {d_2})\in\mathfrak{C}_c$, where $(d_1,d_2)$ is given by (\ref{lem:mapping:eqd1}) for some $\mathbf{r}\in\mathcal{C}_1$ and $\mathfrak{C}_c$ is given in Theorem \ref{theo:network}. After some manipulation, we can show that $\mathbf{R}=(R_1,R_2)$ is given by
\begin{align*}
  R_1=\tfrac{\tau_1}{d_1}=r_1,\quad R_2=\tfrac{\tau_2}{d_2}=\tfrac{\tau_2r_1r_2^*}{\tau_2r_1-\tau_1r_2+r_2^*\tau_1},
\end{align*}
for some $(r_1,r_2)\in\mathcal{C}_1$. If $\frac{1}{c}R_2-(\frac{1}{c}-1)r_2^*\leq 0$, we immediately see that $(r_1,0)\in\mathcal{C}$, since $(r_1,r_2)\in\mathcal{C}_1\subseteq\mathcal{C}$. Hence from Theorem \ref{theo:network}, $\mathbf{R}\in \mathfrak{C}_c$. If $\frac{1}{c}R_2-(\frac{1}{c}-1)r_2^*>0$, we can check that the inequality $\frac{1}{c}R_2-(\frac{1}{c}-1)r_2^*\leq r_2$ holds by plugging in $c=\frac{d_1}{d_2}=\frac{\tau_1R_2}{\tau_2R_1}$ and $R_1=r_1$. Since $(r_1,r_2)\in\mathcal{C}$, from Theorem \ref{theo:network}, we have $\mathbf{R}\in \mathfrak{C}_c$. Furthermore, it is easy to check that $d_1\leq d_2$ and hence $\mathbf{d}\in\mathcal{D}_1^*$.

For the converse, recall from Corollary \ref{newdef} that $\mathcal{D}^*= \{(d_1,d_2):(\frac{\tau_1}{d_1},\frac{\tau_2}{d_2})\in \mathfrak{C}_c, c=\frac{d_1}{d_2}\}$. Plugging in $\mathfrak{C}_c$ from Theorem \ref{theo:network}, we have for $c\leq 1$, $\mathcal{D}_1^*=\{(d_1,d_2):(\frac{\tau_1}{d_1}, [\frac{\tau_2}{d_1}-\frac{d_2-d_1}{d_1}r_2^*]^+)\in\mathcal{C},d_1\leq d_2\}$. For any given $(d_1,d_2)\in\mathcal{D}_1^*$, let
\begin{align}
\label{eq:r1r2}
  r_1=\tfrac{\tau_1}{d_1},\qquad r_2=[\tfrac{\tau_2}{d_1}-\tfrac{d_2-d_1}{d_1}r_2^*]^+.
\end{align}
For the converse, it is sufficient to just consider $d_2\leq\tau_2/r_2^*+d_1$, otherwise $\mathbf{d}$ is dominated by $(d_1,\tau_2/r_2^*+d_1)$. Then $r_2$ above reduces to $r_2=\tau_2/d_1-(d_2-d_1)r_2^*/d_1$. Solving (\ref{eq:r1r2}) for $d_1$ and $d_2$, we obtain  (\ref{lem:mapping:eqd1}). At last, it can be checked that with $d_1$, $d_2$ given by (\ref{lem:mapping:eqd1}), $c=\frac{d_1}{d_2}\leq 1$ implies $\frac{r_2}{r_1}\leq \frac{\tau_2}{\tau_1}$. Therefore for any $\mathbf{d}\in\mathcal{D}_1^*$, there exists some $\mathbf{r}\in\mathcal{C}_1$ such that $\mathbf{d}=G_1(\mathbf{r})$.
\end{IEEEproof}

\begin{remark}
\label{prop:linearity}
The maps $G_i$, $i=1,2,$ given in (\ref{lem:mapping:eqd1}) have the following properties:
\begin{enumerate}
    \item $G_1=G_2$ for $c=1$.
    \item $G_i$ is invertible and non-increasing, i.e. $G_i(\mathbf{r})\leq G_i(\mathbf{r}')$ for $\mathbf{r}\geq\mathbf{r}'$.
    \item $G_i$ is an affine map, meaning that for $i=1$, if $\mathbf{r}\in\mathcal{C}_1$ satisfies $a_1r_1+a_2r_2=1$ for some constants $a_i$, then $\mathbf{d}=G_1(\mathbf{r})\in\mathcal{D}_1^*$ satisfies $a_1'd_1+a_2'd_2=1$, where $a_1'= \frac{1-a_2R_2'} {a_1\tau_1+a_2\tau_2}$ and $a_2'=\frac{a_2R_2'}{a_1\tau_1+a_2\tau_2}$. The case of $i=2$ follows similarly where indices 1 and 2 are swapped.
\end{enumerate}
\end{remark}

\subsection{Convexity of Completion Time Subregions}
\begin{proposition}
\label{prop:convex}
Completion time subregions $\mathcal{D}_i^*$, $i\in\{1,2\}$, defined in (\ref{def:ctrsubregion}) are convex.
\end{proposition}

\begin{figure}[htb]
    \centering
    \subfloat[]{\includegraphics[width=0.35\textwidth]{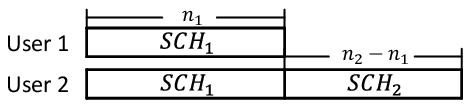}}
    \qquad
    \subfloat[]{\includegraphics[width=0.4\textwidth]{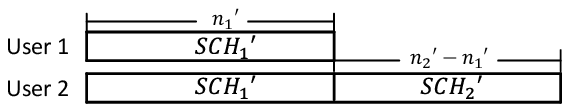}}
    \quad
    \subfloat[]{\includegraphics[width=0.4\textwidth]{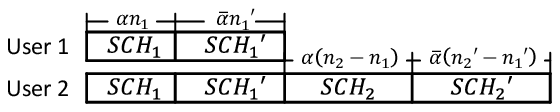}}
    \caption{Transmission schemes achieving $\mathbf{d}$, $\mathbf{d}'$, and $\mathbf{d}''$ in the proof of Proposition \ref{prop:convex}}
    \label{fig:timesharing}
\end{figure}

\begin{IEEEproof}
We prove the proposition for $\mathcal{D}_{1}^*$. The case of $\mathcal{D}_{2}^*$ follows similarly. For $\mathbf{d},\mathbf{d}'\in\mathcal{D}_{1}^*$, we need to show $\mathbf{d}''=\alpha\mathbf{d}+\bar{\alpha}\mathbf{d}'\in\mathcal{D}_{1}^*$, where $0\leq\alpha\leq 1$ and $\bar{\alpha}=1-\alpha$.

Suppose $\mathbf{d}=(d_1,d_2)$ is an achievable completion time pair. In light of Theorem \ref{theo:network}, without loss of generality we  consider a two-phase transmission scheme shown in Fig. \ref{fig:timesharing}(a). In the first $n_1=md_1$ channel uses, coding scheme $SCH_1$ is employed. In the remaining $n_2-n_1=m(d_2-d_1)$ channel uses, coding scheme $SCH_2$ is employed where only user 2 is active. Note that by Theorem \ref{theo:network}, we can view user 2's message as consisting of two independent parts for $n_1$ and $n_2-n_1$ channel uses respectively. Hence the decoding for user 1 and the first part of user 2's message is accomplished after $n_1$ channel uses and the decoding for the second part of user 2's message is accomplished by the end of $n_2-n_1$ channel uses. Similarly for $\mathbf{d}'$, we consider schemes $SCH_1'$ and $SCH_2'$ shown in Fig. \ref{fig:timesharing}(b). Based on the coding schemes for $\mathbf{d}$ and $\mathbf{d}'$, we construct a new coding scheme shown in Fig. \ref{fig:timesharing}(c), where scheme $SCH_1$ is used for the first $\alpha n_1$ channel uses, followed by $SCH_1'$ for $\alpha n_1'$ channel uses, then $SCH_2$ for $\alpha (n_2-n_1)$ channel uses and finally $SCH_2'$ for $\alpha (n_2'-n_1')$ channel uses. Note that here we use four sub-intervals for the time-sharing so that codes in each sub-interval are properly aligned. The completion time achieved in this way for user $i$ is $\frac{\alpha n_i+\bar{\alpha}n_i'}{m}=\alpha d_i+\bar{\alpha}d_i'$. Furthermore, since $d_1\leq  d_2$ and $d_1'\leq d_2'$, we have $\alpha d_1+\bar{\alpha}d_1'\leq\alpha d_2+\bar{\alpha}d_2'$. Therefore $\mathbf{d}''\in\mathcal{D}_1^*$.
\end{IEEEproof}


\begin{figure}[htb]
    \centering
    \includegraphics[width=60mm]{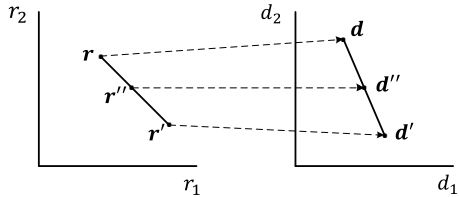}
    \caption{Mapping $G_i$ in Lemma \ref{lem:mapping} preserves linearity}
    \label{fig:mapping}
\end{figure}
The convexity of $\mathcal{D}_i^*$ can also be deduced from the fact that $G_i$ is an affine map, discussed in Remark \ref{prop:linearity}. To see this, suppose there are two achievable completion time pairs $\mathbf{d},\mathbf{d}'\in\mathcal{D}_i^*$ shown in Fig. \ref{fig:mapping}. Then by Lemma \ref{lem:mapping} we can find two achievable rate pairs $\mathbf{r},\mathbf{r}'\in\mathcal{C}_i$ that can be mapped to $\mathbf{d},\mathbf{d}'$ respectively. Due to the convexity of $\mathcal{C}_i$, any point $\mathbf{r}'' \in \overline{\mathbf{r}\mathbf{r}'}$ is also achievable. Since $G_i$ is an affine map, line segment $\overline{\mathbf{d}\mathbf{d}'}$ is mapped from line segment $\overline {\mathbf{r}\mathbf{r}'}$. Therefore by Lemma \ref{lem:mapping} any point $\mathbf{d}''\in \overline{\mathbf{d}\mathbf{d}'}$ is also achievable. However, if $\mathbf{d}\in\mathcal{D}_1^*$ and $\mathbf{d}'\in\mathcal{D}_2^*$, above arguments no longer hold because of the different maps in Lemma \ref{lem:mapping}, namely $G_1$ and $G_2$ respectively.

\subsection{Characterizing the Completion Time Region}
Next, we will characterize the completion time region by considering its two subregions $\mathcal{D}_i^*$, $i\in\{1,2\}$, individually. We will show that the boundary of $\mathcal{D}_i^*$ is given by the image of the boundary of $\mathcal{C}_i$ under the map $G_i$. To this end, we would like to express the capacity region as $\mathcal{C}=\{\mathbf{r}:\mathbf{r}\in \mathbb{R}_2^+, f(\mathbf{r})\leq 0 \}$ for some function $f$. Then the boundary of $\mathcal{C}$ is given by $\{\mathbf{r}: \mathbf{r}\in \mathbb{R}_2^+,f(\mathbf{r})=0 \}$. For a DM-MAC, the characterization of $\mathcal{C}$ given in (\ref{eq:maccapa}), hinges on a time-sharing random variable $Q$, which in general prevents us from obtaining an explicit expression of $f$. Therefore in Theorem \ref{theo:networkctr} that follows, we characterize the completion time region using $f$ without giving its expression. This characterization, although not in an explicit form, reveals some general structural properties of the completion time region. We then illustrate the completion time region for a GMAC, where a closed form for $f$ is available.

\begin{figure}[htb]
    \centering
    \subfloat[Capacity region]{\includegraphics[width=0.23\textwidth]{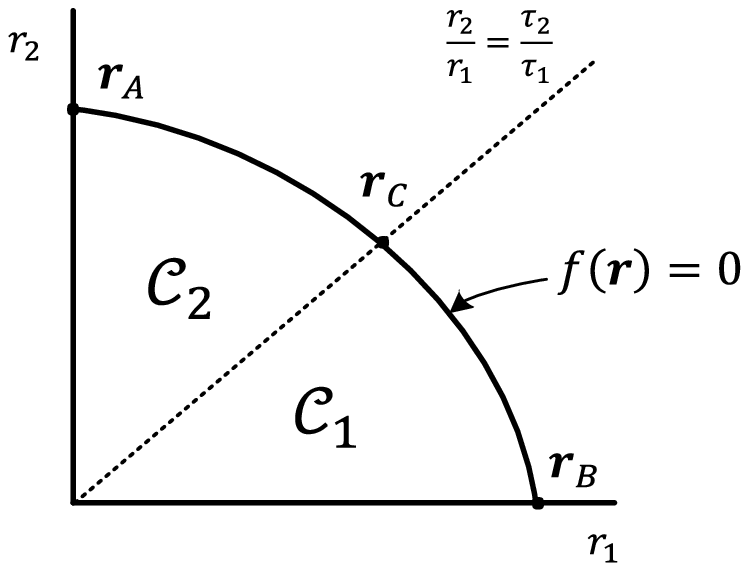}}
    \quad
    \subfloat[Completion time region]{\includegraphics[width=0.21\textwidth]{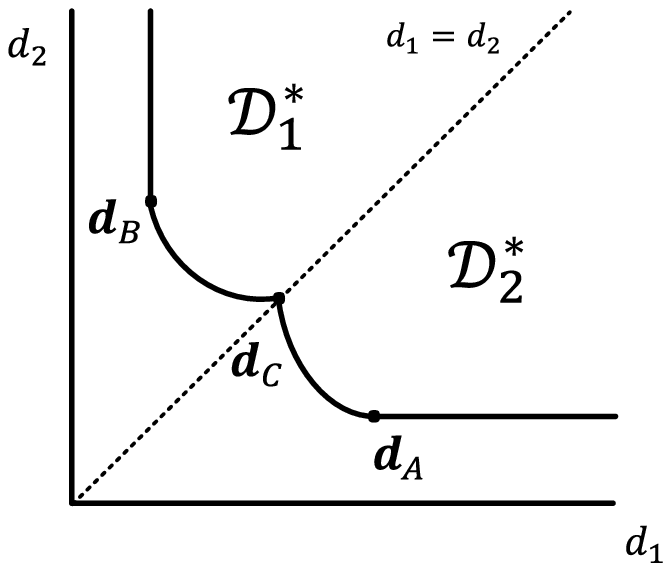}}
    \caption{The capacity region and completion time region of a DM-MAC}
    \label{fig:generalregion}
\end{figure}

Referring to Fig. \ref{fig:generalregion}(a), $\mathbf{r}_A$ and $\mathbf{r}_B$ denote the $r_2$-axis and $r_1$-axis intercepts of the capacity region boundary respectively. $\mathbf{r}_C$ denotes the intersection of line $\frac{r_2}{r_1}=\frac{\tau_2}{\tau_1}$ and the capacity region boundary. Let $\mathbf{d}_A=G_2(\mathbf{r}_A)$, $\mathbf{d}_B=G_1(\mathbf{r}_B)$ and $\mathbf{d}_C= G_1(\mathbf{r}_C) =G_2(\mathbf{r}_C)$, where $G_i$, $i\in\{1,2\}$, is defined in Lemma \ref{lem:mapping}.
\begin{theorem}
\label{theo:networkctr}
For a DM-MAC, the completion time region is given by $\mathcal{D}^*=\mathcal{D}_1^*\bigcup \mathcal{D}_2^*$, where $\mathcal{D}_1^*$ ($\mathcal{D}_2^*$) is an open convex region illustrated in Fig. \ref{fig:generalregion}(b) whose boundary consists of
\begin{enumerate}
\item a vertical (horizontal) ray emanating from $\mathbf{d}_B$ ($\mathbf{d}_A$),
\item a 45-degree ray emanating from $\mathbf{d}_C$,
\item and curve  $\widetilde{\mathbf{d}_B \mathbf{d}_C}$ $\left(\widetilde{\mathbf{d}_A \mathbf{d}_C}\right)$ mapped from $\widetilde{\mathbf{r}_B \mathbf{r}_C}$ $\left(\widetilde{\mathbf{r}_A \mathbf{r}_C}\right)$ using $G_1$ $(G_2)$,
\end{enumerate}
where $\widetilde{\mathbf{r}_B \mathbf{r}_C}=\{\mathbf{r}: \mathbf{r}\in \mathbb{R}_2^+,f(\mathbf{r})=0,\frac{r_2}{r_1}\leq \frac{\tau_2}{\tau_1} \}$, $\widetilde{\mathbf{r}_A \mathbf{r}_C}=\{\mathbf{r}: \mathbf{r}\in \mathbb{R}_2^+,f(\mathbf{r})=0,\frac{r_2}{r_1}\geq \frac{\tau_2}{\tau_1} \}$ and $G_i$, $i\in\{1,2\}$, is given by (\ref{lem:mapping:eqd1}).
\end{theorem}

\begin{IEEEproof}
We prove the theorem for $\mathcal{D}_1^*$. The case of $\mathcal{D}_2^*$ follows similarly.

For the achievability, consider $\widetilde{\mathbf{d}_B \mathbf{d}_C}$ mapped from $\widetilde{\mathbf{r}_B \mathbf{r}_C}$ using $G_1$. First of all any point on $\widetilde{\mathbf{d}_B \mathbf{d}_C}$ is achievable, since $\widetilde{\mathbf{d}_B \mathbf{d}_C}$ is mapped from $\widetilde{\mathbf{r}_B \mathbf{r}_C}$. Secondly any point on the vertical ray emanating from $\mathbf{d}_B$ is achievable. This is because we can use the same codebooks designed for achieving $\mathbf{d}_B$ but decrease the rate of user 2 by using only a subset of the codewords, resulting in the same $d_1$ but a larger $d_2$. For the same reason, any point on the 45-degree ray emanating from $\mathbf{d}_C$ is also achievable. Here we keep the same codebooks for achieving $\mathbf{d}_C$ but decrease both rates by the same amount. Finally, any inner point in the region can be expressed as the convex combination of two points on the boundary and hence is also achievable due to the convexity of $\mathcal{D}_1^*$.

For the converse, we note that for any $\mathbf{r}\in\mathcal{C}_1$, there always exists some point $\mathbf{r}'\in \widetilde{\mathbf{r}_B \mathbf{r}_C}$ such that $\mathbf{r}\leq \mathbf{r}'$, because $\widetilde{\mathbf{r}_B \mathbf{r}_C}$ is the boundary of $\mathcal{C}_1$. It follows directly that for any $\mathbf{d}\in\mathcal{D}_1^*$, where $\mathbf{d}=G_1(\mathbf{r})$ for some $\mathbf{r}\in\mathcal{C}_1$, there always exists some $\mathbf{d}'\in\widetilde{\mathbf{d}_B \mathbf{d}_C}$, where $\mathbf{d}'=G_1(\mathbf{r}')$ for some $\mathbf{r}'\in \widetilde{\mathbf{r}_B \mathbf{r}_C}$, such that $\mathbf{d}'\leq \mathbf{d}$, because $G_1$ is a non-increasing function of $\mathbf{r}$. The converse proof is complete by noting that $d_1\leq d_2$ by the definition of $\mathcal{D}_1^*$.
\end{IEEEproof}

As an illustration, we next explicitly compute the completion time region of a GMAC given by
\begin{align}
\label{channelGMAC}
    Y=X_1+X_2+Z,
\end{align}
where $Z\sim\mathcal{N}(0,1)$ is the i.i.d. Gaussian noise process. We assume expected per symbol power constraints on the input distribution: $E[X_i^2]\leq P_i$, $i\in\{1,2\}$.
\begin{remark}
In the expected per symbol power constraint, the expectation is over the message set and the maximum transmission power is held fixed across time. This restriction simplifies the analysis, leading to a closed form characterization of the completion time region. With the more general expected block power constraint, the maximum transmission power is allowed to vary in different phases of operation. This variation complicates the analysis and is presented in Section VI.
\end{remark}

\begin{figure}[htb]
    \centering
    \includegraphics[width=85mm]{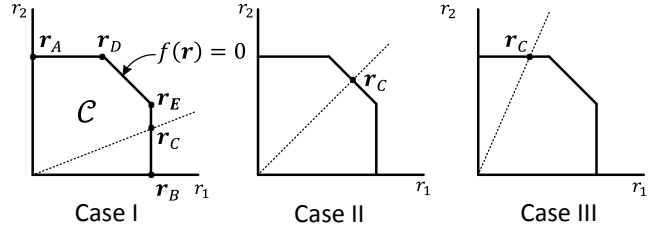}
    \caption{Three cases of the location where line $\frac{r_2}{r_1}=\frac{\tau_2}{\tau_1}$ intersects the boundary of GMAC capacity region}
    \label{fig:MACThreeCase}
\end{figure}

For a GMAC, we have  $r_i^*=\gamma(P_i)$ for $i\in\{1,2\}$, where $\gamma(x)=\frac{1}{2}\log(1+x)$,  and
\begin{align}
\label{eq:GMACCapa}
    \mathcal{C}= \left\{(r_1,r_2):  0\leq \sum_{i\in \Omega}r_i\leq \gamma(\sum_{i\in \Omega}P_i),\ \Omega\subseteq \{1,2\} \right\}=\left\{\mathbf{r}:\mathbf{r}\in \mathbb{R}_2^+, f(\mathbf{r})\leq 0 \right\} ,
\end{align}
where $f$ is a piece-wise linear function given by
\begin{align}
\label{eq:macf}
  f(\mathbf{r})=\begin{cases}
                 r_2-\gamma(P_2),\quad r_1\leq\gamma(\frac{P_1}{1+P_2})\\
		 r_1+r_2-\gamma(P_1+P_2), \quad \gamma(\frac{P_i}{1+P_j})\leq r_i \leq \gamma(P_i),\ i,j\in\{1,2\},\ i\neq j\\
		 r_1-\gamma(P_1), \quad r_2\leq\gamma(\frac{P_2}{1+P_1})
                \end{cases}.
\end{align}
The expression of $\mathcal{D}^*$ of a GMAC depends on where line $\frac{r_2}{r_1}=\frac{\tau_2}{\tau_1}$ intersects the capacity region boundary. Accordingly we define three cases as shown in Fig. \ref{fig:MACThreeCase}.
\begin{align}
\label{cases}
\textrm{Case I}: \tfrac{\tau_2}{\tau_1}\leq q_1, \quad \textrm{Case II}: q_1<\tfrac{\tau_2}{\tau_1}< q_2,\quad  \textrm{Case III}: \tfrac{\tau_2}{\tau_1}\geq q_2,
\end{align}
where $q_1= \frac{\gamma(P_1+P_2)-\gamma(P_1)} {\gamma(P_1)}$ and $q_2=\frac{\gamma(P_2)}{\gamma(P_1+P_2)-\gamma(P_2)}$.

\begin{figure}[htb]
    \centering
    \includegraphics[width=100mm]{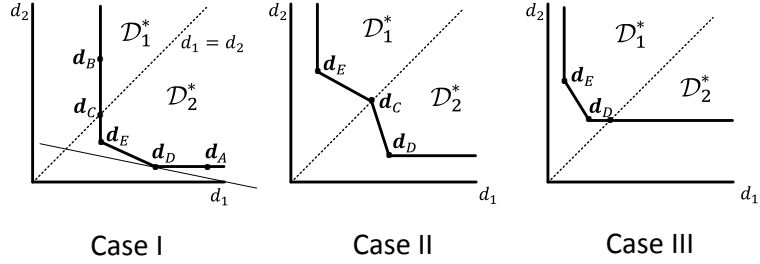}
    \caption{The completion time region of a GMAC}
    \label{fig:MACCTRegion}
\end{figure}

\begin{corollary}
\label{theo:GMACCTR}
The completion time region of the GMAC in (\ref{channelGMAC}), illustrated in Fig. \ref{fig:MACCTRegion}, is given by
\begin{enumerate}
\item Case I
\begin{align}
\label{CTRproofeq}
    \mD^*=\left\{\begin{aligned}
    &(d_1,d_2)\in \mathbb{R}^2_+:\ d_1\geq \tfrac{\tau_1}{\gamma(P_1)},\ d_2\geq \tfrac{\tau_2}{\gamma(P_2)}, \\
    &\gamma(P_1)d_1+[\gamma(P_1+P_2)-\gamma(P_1)]d_2\geq \tau_1+\tau_2
    \end{aligned}\right\}
\end{align}
\item Case II
\begin{align*}
    \mD^*=\left\{\begin{aligned}
    &(d_1,d_2)\in \mathbb{R}^2_+:\ d_1\geq \tfrac{\tau_1}{\gamma(P_1)},\ d_2\geq \tfrac{\tau_2}{\gamma(P_2)}, \\
    &[\gamma(P_1+P_2)-\gamma(P_2)]d_1+\gamma(P_2)d_2\geq \tau_1+\tau_2,\\
    &\gamma(P_1)d_1+[\gamma(P_1+P_2)-\gamma(P_1)]d_2\geq \tau_1+\tau_2
    \end{aligned}\right\}
\end{align*}
\item Case III
\begin{align*}
    \mathcal{D}^*=\left\{\begin{aligned}
    &(d_1,d_2)\in \mathbb{R}^2_+:\ d_1\geq \tfrac{\tau_1}{\gamma(P_1)},\ d_2\geq \tfrac{\tau_2}{\gamma(P_2)}, \\
    &[\gamma(P_1+P_2)-\gamma(P_2)]d_1+\gamma(P_2)d_2\geq \tau_1+\tau_2
    \end{aligned}\right\}
\end{align*}
\end{enumerate}
\end{corollary}

\begin{IEEEproof}
Here we prove the theorem for Case I. The others follow similarly. We use Theorem \ref{theo:networkctr} with $f$ given by (\ref{eq:macf}). Specifically for $\mathcal{D}_1^*$, $\widetilde{\mathbf{r}_B \mathbf{r}_C}$ in Fig. \ref{fig:generalregion}(a) reduces to a line segment
\begin{align*}
\overline{\mathbf{r}_B\mathbf{r}_C}=\{(r_1,r_2):r_1=\gamma(P_1), 0\leq r_2\leq \tfrac{\tau_2}{\tau_1}\gamma(P_1)\}
\end{align*}
in Fig. \ref{fig:MACThreeCase}, which is then mapped to
\begin{align*}
\overline {\mathbf{d}_B \mathbf{d}_C}=\{(d_1,d_2):d_1=\tfrac{\tau_1}{\gamma(P_1)},\tfrac{\tau_1}{\gamma(P_1)}\leq d_2\leq \tfrac{\tau_1}{\gamma(P_1)}+\tfrac{\tau_2}{\gamma(P_2)}\}
\end{align*}
in Fig. \ref{fig:MACCTRegion} Case I, using $G_1$ in (\ref{lem:mapping:eqd1}). Similarly for $\mathcal{D}_2^*$, $\widetilde{\mathbf{r}_A \mathbf{r}_C}$ in Fig. \ref{fig:generalregion}(a) reduces to three line segments
\begin{align*}
\overline {\mathbf{r}_A \mathbf{r}_D}&=\{(r_1,r_2):0\leq r_1\leq \gamma(\tfrac{P_1}{1+P_2}),r_2=\gamma(P_2)\} \\
\overline{\mathbf{r}_D\mathbf{r}_E}&=\{(r_1,r_2):r_1+r_2-\gamma(P_1+P_2)=0, r_1 \leq \gamma(P_1), r_2 \leq \gamma(P_2)\}\\
\overline{\mathbf{r}_E\mathbf{r}_C}&=\{(r_1,r_2):r_1=\gamma(P_1), \tfrac{\tau_2}{\tau_1}\gamma(P_1)\leq r_2\leq\gamma(\tfrac{P_2}{1+P_1})\}
\end{align*}
in Fig. \ref{fig:MACThreeCase}, which are then respectively mapped to
\begin{align*}
\overline {\mathbf{d}_A \mathbf{d}_D}&=\{(d_1,d_2):\tfrac{1}{\gamma(P_1)}[\tau_1+\tau_2-(\gamma(P_1+P_2)-\gamma(P_1))\tfrac{\tau_2}{\gamma(P_2)}]\leq d_1\leq \tfrac{\tau_1}{\gamma(P_1)}+\tfrac{\tau_2}{\gamma(P_2)},d_2=\tfrac{\tau_2}{\gamma(P_2)}\}\\
\overline {\mathbf{d}_D \mathbf{d}_E}&=\{(d_1,d_2): d_1\geq \tfrac{\tau_1}{\gamma(P_1)},d_2\geq \tfrac{\tau_2}{\gamma(P_2)},\gamma(P_1)d_1+[\gamma(P_1+P_2)-\gamma(P_1)]d_2= \tau_1+\tau_2\}\\
\overline{\mathbf{d}_E\mathbf{d}_C}&=\{(d_1,d_2):d_1=\tfrac{\tau_1}{\gamma(P_1)},\tfrac{\tau_2}{\gamma(P_1+P_2)-\gamma(P_1)}\leq d_2\leq \tfrac{\tau_1}{\gamma(P_1)}\}
\end{align*}
in Fig. \ref{fig:MACCTRegion} Case I, using $G_2$ in (\ref{lem:mapping:eqd1}). Together with the horizontal ray emanating from $\mathbf{d}_A=(\frac{\tau_1}{\gamma(P_1)}+\frac{\tau_2}{\gamma(P_2)},\frac{\tau_2}{\gamma(P_2)})$ and the vertical ray emanating from $\mathbf{d}_B=(\frac{\tau_1}{\gamma(P_1)},\frac{\tau_1}{\gamma(P_1)}+\frac{\tau_2}{\gamma(P_2)})$, we obtain the boundary of $\mathcal{D}^*$ which gives rise to the expression of $\mathcal{D}^*$ in (\ref{CTRproofeq}).
\end{IEEEproof}

\begin{remark}
In Fig. \ref{fig:MACCTRegion} Case II, the slopes of line $\overline{\mathbf{d}_D\mathbf{d}_C}$ and $\overline {\mathbf{d}_E \mathbf{d}_C}$ are given by $-\frac{\gamma(P_1)}{\gamma(P_1+P_2)-\gamma(P_1)}$ and $-\frac{\gamma(P_1+P_2)-\gamma(P_2)}{\gamma(P_2)}$ respectively. Due to the concavity of the logarithm function, the former is smaller than the latter. Therefore the completion time region of a GMAC is not convex in Case II. This shows that while subregion $\mathcal{D}_1^*$ and $\mathcal{D}_2^*$ are convex, the whole completion time region $\mathcal{D}^*$ is not convex in general.
\end{remark}

\section{Completion Time Region of a Gaussian Broadcast Channel and a Gaussian Interference Channel}
The steps of deriving the completion time region of a DM-MAC detailed in the previous section can be extended to the broadcast channel and the interference channel as well. While the general formulation holds for discrete memoryless channels, for ease of exposure, we focus on Gaussian channels only, which permit computable completion time regions. Since the arguments in this section mostly parallel those for a DM-MAC, to highlight to difference, we will only emphasize the parts pertaining to a GBC and a GIC. For notational economy, we use the same notation for similar quantities. For example, $\mathcal{C}$ refers to the capacity region. Whether it is the capacity region of a GBC or a GIC, will be understood from the context.

\subsection{Completion Time Region of a Gaussian Broadcast Channel}
Consider a Gaussian broadcast channel
\begin{align}
\label{channelGBC}
    Y_i = \sqrt{h_i}X + Z_i,\quad i\in\{1,2\},
\end{align}
where $Z_i\sim\mathcal{N}(0,1)$ is the i.i.d. Gaussian noise process and we assume an expected per symbol power constraint: $E[X^2]\leq P$. Without loss of generality, we assume $h_1\geq h_2$. Hence the capacity region of a GBC is given by
\begin{align}
\label{eq:ratebc}
    \mathcal{C}= \left\{(r_1,r_2):  0\leq r_1 \leq \gamma(h_1P_1),\ 0\leq r_2\leq \gamma(h_2P) - \gamma(h_2P_1),\ 0\leq P_1\leq P  \right\}.
\end{align}

From Section III-A, the first step of deriving the completion time of a DM-MAC is to obtain the $c$-constrained capacity region of a DM-MAC. The theorem below represents a similar result for the $c$-constrained capacity region of a GBC.
\begin{theorem}
\label{coro:bcconcapacity}
The $c$-constrained capacity region $\mathfrak{C}_c$ of the GBC in (\ref{channelGBC}) is the set of all non-negative $(R_1,R_2)$ satisfying
\begin{enumerate}
\item
$(R_1,[\frac{1}{c}R_2-(\frac{1}{c}-1)\gamma(P)]^+)\in\mathcal{C}$ for $c\leq 1$,
\item
$([cR_1-(c-1)\gamma(P)]^+,R_2)\in\mathcal{C}$ for $c\geq 1$,
\end{enumerate}
where $\mathcal{C}$ is given by (\ref{eq:ratebc}).
\end{theorem}
\begin{IEEEproof}
The proof parallels that of Theorem \ref{theo:network} and is omitted.
\end{IEEEproof}

The next step, discussed in Section III-B and III-C, is to use the obtained $c$-constrained capacity region to derive a mapping between the completion time region and the standard capacity region as in Lemma \ref{lem:mapping}, and argue the convexity of the completion time subregions as in Proposition \ref{prop:convex}. One can easily observe that these steps are channel independent. Therefore, extensions of Lemma \ref{lem:mapping} and Proposition \ref{prop:convex} for a GBC follow immediately. The final step, discussed in Section III-D, is to map the boundary of the standard capacity region of a DM-MAC into that of the completion time region. Since Theorem \ref{theo:networkctr} depends on the specific channel only through the capacity region boundary function $f$, it can also be extended to a GBC. We next present an explicit characterization of the completion time region of a GBC in Corollary \ref{theo:CTRGBC}.

We first rewrite the capacity region of a GBC given by (\ref{eq:ratebc}) as $\mathcal{C}=\left\{\mathbf{r}:\mathbf{r}\in \mathbb{R}_2^+, f(\mathbf{r})\leq 0 \right\}$, where
\begin{align}
\label{GBCcapaboundary}
f(\mathbf{r})=r_2+\gamma\left(\left(2^{2r_1}-1\right)h_2/h_1\right)-\gamma(h_2P).
\end{align}
As in Section III-D, $\mathbf{r}_C$ denotes the intersection of line $\frac{r_2}{r_1}=\frac{\tau_2}{\tau_1}$ and the boundary of $\mathcal{C}$. For any given $\frac{r_2}{r_1}$, $\mathbf{r}_C=(r_{1,C}, \frac{\tau_2}{\tau_1}r_{1,C})$, $r_{1,C}=\gamma(h_1P_{1,C})$, can be solved numerically for some $P_{1,C}\in[0, P]$.

\begin{figure}[htb]
  \centering
  \psfrag{r1}{$r_1$}
  \psfrag{r2}{$r_2$}
  \psfrag{C1}{$\mathcal{C}_1$}
  \psfrag{C2}{$\mathcal{C}_2$}
  \psfrag{ra}[c][c][0.8][0]{$\mathbf{r}_A$}
  \psfrag{rb}[c][c][0.8][0]{$\mathbf{r}_B$}
  \psfrag{rc}[c][c][0.8][0]{$\mathbf{r}_C$}
  \subfloat[GBC capacity region]{\includegraphics[width=55mm]{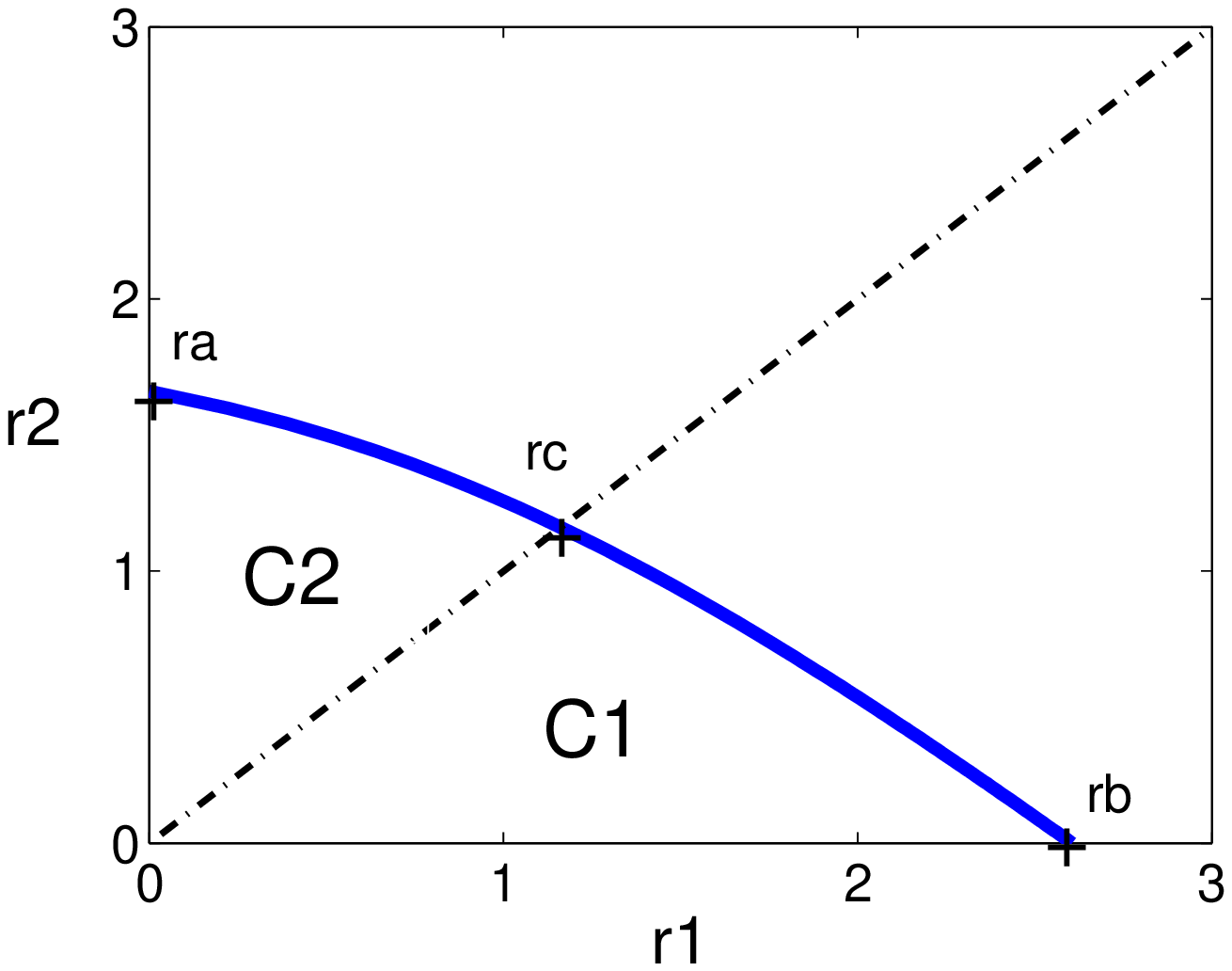}}
  \quad
  \psfrag{d1}{$d_1$}
  \psfrag{d2}{$d_2$}
  \psfrag{d1star}{$\mathcal{D}_1^*$}
  \psfrag{da}[c][c][0.8][0]{$\mathbf{d}_A$}
  \psfrag{db}[c][c][0.8][0]{$\mathbf{d}_B$}
  \psfrag{dc}[c][c][0.8][0]{$\mathbf{d}_C$}
  \psfrag{s1}[c][c][0.8][0]{$s_1$}
  \psfrag{s2}[c][c][0.8][0]{$s_2$}
  \psfrag{d2star}{$\mathcal{D}_2^*$}
  \subfloat[GBC completion time region]{\includegraphics[width=55mm]{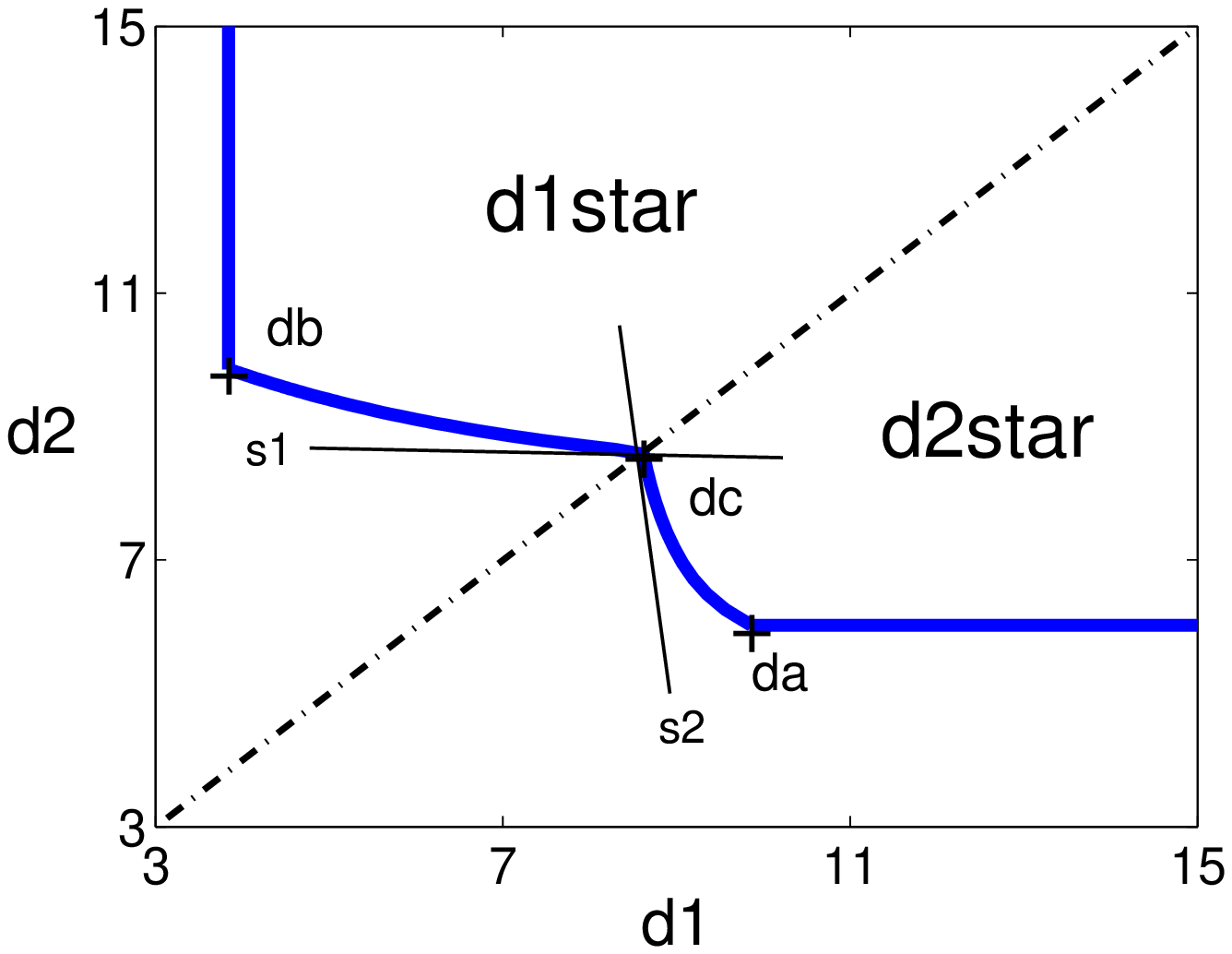}}
  \caption{GBC with $\tau_1=\tau_2=10,\ h_1=4,\ h_2=1,\ P=9$}
  \label{fig:BCCTR}
\end{figure}

\begin{corollary}
\label{theo:CTRGBC}
The completion time region of the GBC in (\ref{channelGBC}) is given by $\mathcal{D}^* = \mathcal{D}_1^* \bigcup\mathcal{D}_2^*$, where
\begin{align*}
    \mathcal{D}_1^*=\left\{
    \begin{aligned}
        &(d_1,d_2)\in \mathbb{R}^2_+:\ \textrm{for}\ P_1\in[P_{1,C},P]  \\
        &d_1\geq \tfrac{\tau_1}{\gamma(h_1P_1)},\ d_2\geq d_1,\\
        &d_2\geq \tfrac{\tau_2}{\gamma(h_2P)} + \tfrac{\gamma(h_2P_1)\tau_1}{\gamma(h_2P)\gamma(h_1P_1)}
     \end{aligned}
     \right\}, \
    \mathcal{D}_2^*=\left\{
    \begin{aligned}
        &(d_1,d_2)\in \mathbb{R}^2_+: \ \textrm{for}\ P_1\in[0,P_{1,C}] \\
        &d_1\geq \tfrac{\tau_1}{\gamma(h_1P)} + \tfrac{[\gamma(h_1P)-\gamma(h_1P_1)]\tau_2} {\gamma(h_1P)[\gamma(h_2P)-\gamma(h_2P_1)]},\\
        &d_2\geq \tfrac{\tau_2}{\gamma(h_2P)-\gamma(h_2P_1)},\ d_1\geq d_2,
     \end{aligned}
     \right\}.
\end{align*}
\end{corollary}
\begin{IEEEproof}
The proof is similar to that of Corollary \ref{theo:GMACCTR}, where the boundary of $\mathcal{D}^*$ is mapped from that of $\mathcal{C}$ in (\ref{channelGBC}) and $f(\mathbf{r})$ given by (\ref{GBCcapaboundary}) is used to obtain the explicit expression of $\mathcal{D}^*$.
\end{IEEEproof}

An example of the completion time region of a GBC is shown in Fig. \ref{fig:BCCTR}(b) for $\tau_1=\tau_2=10,\ h_1=4,\ h_2=1,\ P=9$ leading to $P_{1,C}=1$. Clearly $\mathcal{D}^*$ is not convex for this choice of parameters. In fact, this holds for an arbitrary GBC: $\mathcal{D}^*$ of a GBC is not convex regardless of the channel parameters. The proof will be given in Proposition \ref{prop:BCnoconvex} in Section V in the context of utility optimization using completion time region.

\subsection{Completion Time Region of a Gaussian Interference Channel}
Consider a Gaussian interference channel
\begin{align}
\label{channelGIC}
    Y_1 = X_1 + \sqrt{b}X_2 + Z_1,\quad Y_2 = \sqrt{a}X_1 + X_2 + Z_2,
\end{align}
where $Z_i\sim\mathcal{N}(0,1)$, $i\in\{1,2\}$, is the i.i.d. Gaussian noise process and inputs are subject to expected per symbol power constraints: $E[X_i^2]\leq P_i$. Depending on the values of the non-negative interfering link gains $a,b$, the GIC can be categorized into (\cite{Etkin}) the \textit{very strong} interference regime if $a\geq 1+P_2$, $b\geq 1+P_1$; the \textit{strong} interference regime if $1\leq a< 1+P_2$,  $1\leq b< 1+P_1$; the \textit{weak} interference regime if $a< 1$, $b<1$; the \textit{mixed} interference regime if one interference link is strong and the other is weak. In the following, the exact completion time regions will be established for the very strong and strong interference regimes. For the weak and mixed interference regimes, inner and outer bounds of the completion time region will be obtained.

\begin{proposition}
\label{Prop:verystrong}
The completion time region of the GIC in (\ref{channelGIC}) in the very strong interference regime is $\mathcal{D}^*=\{(d_1,d_2):d_i\geq \frac{\tau_i}{\gamma(P_i)},\ i=1,2\}$.
\end{proposition}

\begin{IEEEproof}
In the very strong interference regime, the capacity region of a GIC is given by $\mathcal{C}=\{(r_1,r_2): \ 0\leq r_i\leq \gamma(P_i), i=1,2 \}$. In this regime, the GIC can be decoupled into two point-to-point channels. As a result, the $c$-constrained and standard capacity regions coincide, i.e. $\mathfrak{C}_c=\mathcal{C}$ for all $c$. In this case $\mathcal{D}^*$ can be directly obtained by Corollary \ref{newdef}.
\end{IEEEproof}

\begin{proposition}
\label{Prop:strong}
The completion time region of the GIC in (\ref{channelGIC}) in the strong interference regime is the same as that of a GMAC given in Corollary \ref{theo:GMACCTR}, except that the term $\gamma(P_1+P_2)$ is replaced by $\min\{\gamma(P_1+bP_2),\gamma(aP_1+P_2)\}$.
\end{proposition}

\begin{IEEEproof}
In the strong interference regime, since the capacity achieving scheme requires each receiver to decode both the desired signal and the interference, the capacity region is the same as that of the compound MAC formed at the two receivers. The capacity region expression is equivalent to that of a GMAC except that the sum rate term $\gamma(P_1+P_2)$ is replaced by $\min\{\gamma(P_1+bP_2),\gamma(aP_1+P_2)\}$. Consequently, $\mathfrak{C}_c$ and $\mathcal{D}^*$ of a GIC in the strong interference regime differ from those of a GMAC by the sum rate term.
\end{IEEEproof}

Even though the capacity region $\mathcal{C}$ of a GIC in the weak or mixed interference regime is still unknown, it was characterized in \cite{Etkin} up to a one-bit gap. Following similar arguments as in Theorem \ref{theo:network}, inner and outer bounds of $\mathcal{C}$ provided in \cite{Etkin} can be used to obtain those of the $c$-constrained capacity region $\mathfrak{C}_c$. Let $\mathcal{R}_W$ and $\overline{\mathcal{R}_W}$ (similarly $\mathcal{R}_M$, $\overline{\mathcal{R}_M}$) denote respectively the inner and outer bounds for the weak (mixed) interference regime given in \cite{Etkin}.

\begin{corollary}
\label{coro:iccapacity}
Let $\mathcal{Q}$ be the set of all non-negative $(R_1,R_2)$ satisfying
\begin{enumerate}
\item
$(R_1,[\tfrac{1}{c}R_2-(\tfrac{1}{c}-1)\gamma(P_2)]^+)\in\mathcal{O}$ for $c\leq 1$,
\item
$([cR_1-(c-1)\gamma(P_1)]^+,R_2)\in\mathcal{O}$ for $c\geq 1$,
\end{enumerate}
for some region $\mathcal{O}\subset\mathbb{R}_2^+$.

For the GIC in (\ref{channelGIC}) in the weak interference regime, if $\mathcal{O}=\mathcal{R}_W$ ($\mathcal{O}=\overline{\mathcal{R}_W}$), then $\mathcal{Q}\subseteq\mathfrak{C}_c$ ($\mathcal{Q}\supseteq\mathfrak{C}_c$). For a GIC in the mixed interference regime, if $\mathcal{O}=\mathcal{R}_M$ ($\mathcal{O}=\overline{\mathcal{R}_M}$), then $\mathcal{Q}\subseteq\mathfrak{C}_c$ ($\mathcal{Q}\supseteq\mathfrak{C}_c$).
\end{corollary}

\begin{IEEEproof}
The proof is similar to the converse part of the proof of Theorem \ref{theo:network} and is omitted.
\end{IEEEproof}

\begin{figure}[htb]
  \centering
  \psfrag{r1}[c][c][0.8][0]{$r_1$}
  \psfrag{r2}[c][c][0.8][0]{$r_2$}
  \subfloat[Inner bound $\mathcal{R}_W$ (solid) and outer bound $\overline{\mathcal{R}_W}$ (dashed) of the capacity region]{\includegraphics[width=49mm]{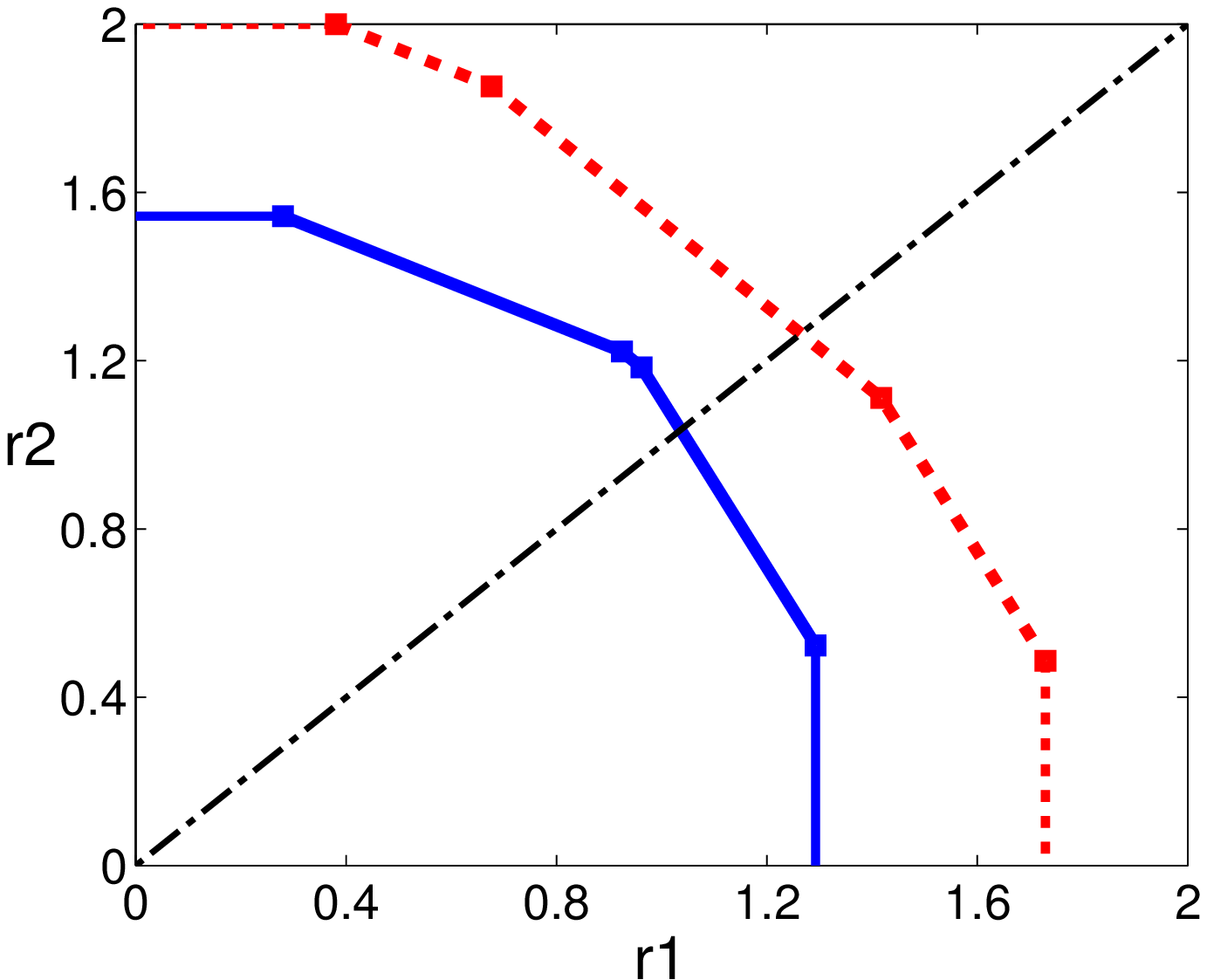}}
  \qquad
  \psfrag{d1}[c][c][0.8][0]{$d_1$}
  \psfrag{d2}[c][c][0.8][0]{$d_2$}
  \subfloat[Inner bound $\mathcal{D}$ (solid) and outer bound $\overline{\mathcal{D}}$ (dashed) of the completion time region]{\includegraphics[width=50mm]{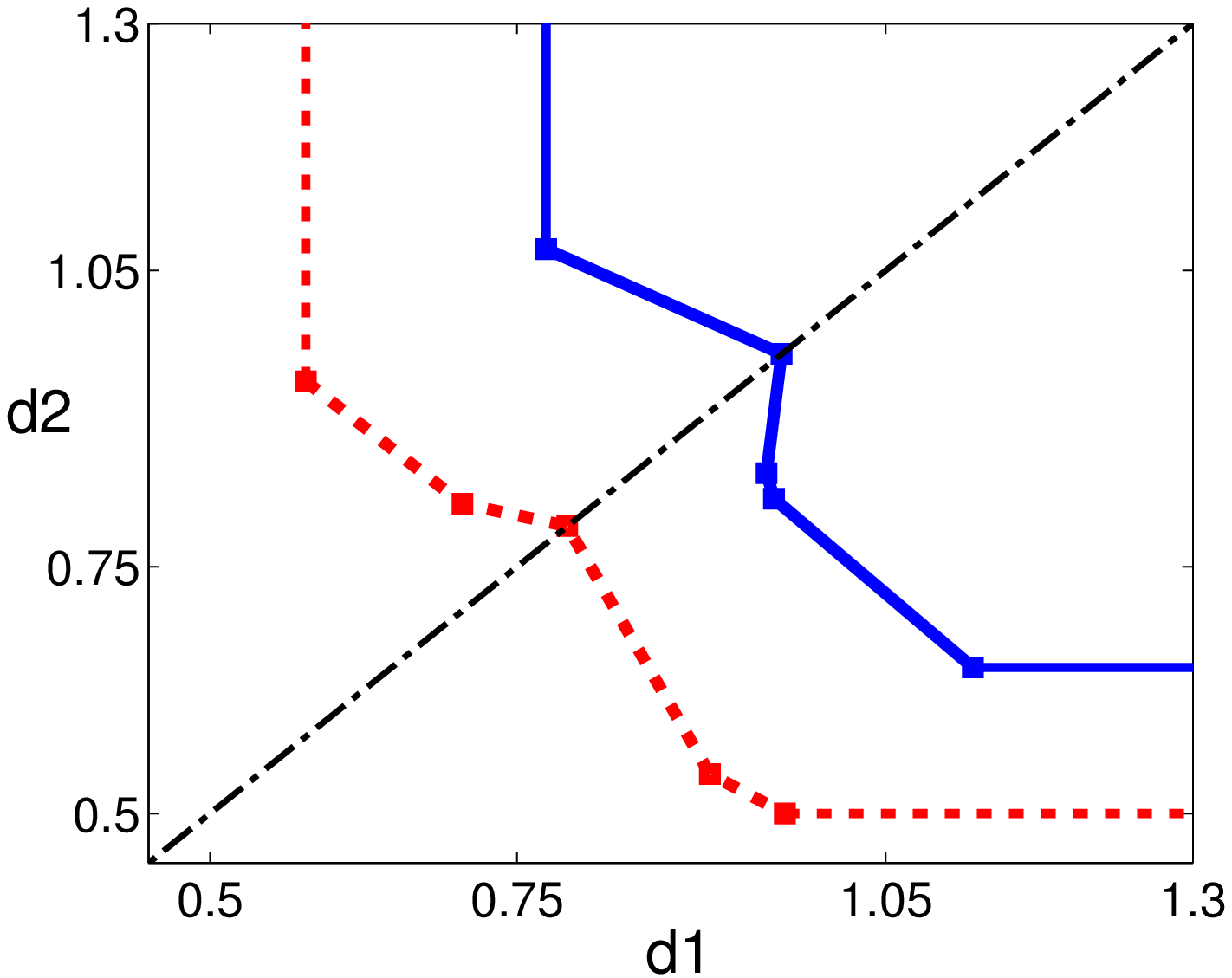}}
  \caption{GIC in the weak interference regime with $P_1=10$, $P_2=15$, $a=0.64$, $b=0.36$, $\tau_1=\tau_2=1$}
  \label{fig:GICWeak}
\end{figure}
Finally arguments similar to Theorem \ref{theo:networkctr} can be used to obtain inner and outer bounds of the completion time region for a GIC in the weak and mixed interference regime. Note that the boundary of $\mathcal{R}_W$, similarly $\overline{\mathcal{R}_W}$, $\mathcal{R}_M$ and $\overline{\mathcal{R}_M}$), can be expressed as $\{\mathbf{r}:\mathbf{r} \in\mathbb{R}_2^+, f(\mathbf{r})=0\}$ where $f$ is a piece-wise linear function. Similar to Corollary \ref{theo:GMACCTR} for a GMAC, we can map $\mathcal{R}_W$ boundary, using $G_i$, $i\in\{1,2\}$, to that of an achievable completion time region $\mathcal{D}$. The exact expression of $\mathcal{D}$ depends on where line $\frac{r_2}{r_1}=\frac{\tau_2}{\tau_1}$ intersects $\mathcal{R}_W$ boundary. As an illustration, the completion time region inner bound $\mathcal{D}$ and outer bound $\overline{\mathcal{D}}$ for a GIC in the weak interference regime with $P_1=10$, $P_2=15$, $a=0.64$, $b=0.36$, $\tau_1=\tau_2=1$ are shown in Fig. \ref{fig:GICWeak}. As in a GMAC and a GBC, both $\mathcal{D}$ and $\overline{\mathcal{D}}$ are non-convex.

\section{Utility Optimization using the Completion Time Region}
Network design is often driven by the goal of optimizing a certain utility, which for example can be a function of users' rates or delays. Equipped with the completion time region, one could seek optimization of a utility that is a function of users' completion times. Because of the way completion time is formulated and $\mathcal{D}^*$ is derived, the information-theoretic optimality of the resultant solution is always guaranteed. As an illustration, one such optimization is sought in this section: minimizing the weighted sum completion time:
\begin{align}
    &\textrm{minimize}\quad\, d_s=wd_1+\bar{w}d_2 \label{eq:optoverall}\\
    &\textrm{subject to}\quad (d_1,d_2)\in\mathcal{D}^* \notag
\end{align}
where $\bar{w}=1-w$, $w\in[0,1]$, and $\mathcal{D}^*$ is the completion time region. This problem can be of practical interest when the network is designed to maximize total user satisfaction, which in this paper is modeled as a simple linear function of the transmission delay --- completion time --- each user experiences for ease of exposure. For an illustration, we next derive analytic solutions of this problem for a GMAC in Section V-A and a GBC in Section V-B.

\subsection{Weighted Sum Completion Time Minimization in a GMAC}
In this subsection, we consider the weighted sum completion time minimization problem (\ref{eq:optoverall}) where $\mathcal{D}^*$  is the completion time region of a GMAC given in Corollary \ref{theo:GMACCTR}. Geometrically, the minimizer will be given by some boundary point of $\mathcal{D}^*$ such that the supporting line of $\mathcal{D}^*$ at that point has slope $s=\frac{w}{w-1}$. The simple analytic form of $\mathcal{D}^*$ of a GMAC makes this approach easy to follow. Before we present the solution, let us consider the following notations.

For Case I, II, III defined in (\ref{cases}), let $\mathbf{d}_D$ and $\mathbf{d}_E$ denote the corners of $\mathcal{D}^*$ shown in Fig. \ref{fig:MACCTRegion}. Then we have
\begin{enumerate}
\item Case I: $\mathbf{d}_D=( \varphi_1, \ \tfrac{\tau_2}{\gamma(P_2)} )$, $\mathbf{d}_E=( \tfrac{\tau_1}{\gamma(P_1)},\ \tfrac{\tau_2}{\gamma(P_1+P_2)-\gamma(P_1)} )$
\item Case II: $\mathbf{d}_D=( \varphi_1, \ \tfrac{\tau_2}{\gamma(P_2)} )$, $\mathbf{d}_E=( \tfrac{\tau_1}{\gamma(P_1)},\ \varphi_2 )$
\item Case III: $\mathbf{d}_D=( \tfrac{\tau_1}{\gamma(P_1+P_2)-\gamma(P_2)} ,\ \tfrac{\tau_2}{\gamma(P_2)})$, $\mathbf{d}_E=( \tfrac{\tau_1}{\gamma(P_1)},\ \varphi_2)$
\end{enumerate}
where
\begin{align*}
  \varphi_1 &= \tfrac{1}{\gamma(P_1)}\left[ \tau_1+\tau_2 -(\gamma(P_1+P_2)-\gamma(P_1))\tfrac{\tau_2}{\gamma(P_2)} \right],\\
  \varphi_2 &= \tfrac{1}{\gamma(P_2)}\left[ \tau_1+\tau_2 -(\gamma(P_1+P_2)-\gamma(P_2))\tfrac{\tau_1}{\gamma(P_1)} \right].
\end{align*}

\begin{theorem}
\label{theo:macweightedmini}
For the GMAC in (\ref{channelGMAC}), the solution of the weighted sum completion time minimization problem (\ref{eq:optoverall}) is given by Table \ref{table:macweightedmini}, where $w_1=\tfrac{\gamma(P_1)}{\gamma(P_1+P_2)}$, $w_2=\frac{\tau_1}{\tau_1+\tau_2}$, $w_3= \tfrac{\gamma(P_1+P_2)-\gamma(P_2)}{\gamma(P_1+P_2)}$.
\end{theorem}

\begin{center}
\begin{table}[htbp]
\caption{Solution to the weighted sum completion time minimization problem for a GMAC}
\label{table:macweightedmini}
\hfill{}
\begin{tabular}{ccc}
\hline\hline
     Case I  &  Case II &  Case III \\
\hline
$\mathbf{d}_D$, $w\in[0,w_1]$ & $\mathbf{d}_D$, $w\in[0,w_2]$ & $\mathbf{d}_D$, $w\in[0,w_3]$ \\\hline
$\mathbf{d}_E$, $w\in(w_1,1]$ & $\mathbf{d}_E$, $w\in(w_2,1]$ & $\mathbf{d}_E$, $w\in(w_3,1]$ \\
\hline
\end{tabular}
\hfill{}
\end{table}
\end{center}

\begin{IEEEproof}
Consider Case I first. The slope of line $\overline{\mathbf{d}_D\mathbf{d}_E}$ is $s_{de}= -\frac{\gamma(P_1)}{\gamma(P_1+P_2) -\gamma(P_1)}$. Any line with negative slope larger than $s_{de}$ supports $\mathcal{D}^*$ at point $\mathbf{d}_D$. Hence $\mathbf{d}_D$ minimizes $d_s$ for $w\leq w_1=\frac{s_{de}}{s_{de} -1}$. Reversely, for $w> w_1$, $\mathbf{d}_E$ becomes the supporting point of $\mathcal{D}^*$ and hence solves problem (\ref{eq:optoverall}). Similarly for Case II, we have $s_{de} = -\frac{\tau_1}{\tau_2}$ and for Case III, $s_{de}= -\frac{\gamma(P_1+P_2) -\gamma(P_2)}{\gamma(P_2)}$. The remaining arguments are the similar as in Case I.
\end{IEEEproof}

In \cite{Rai}, the authors solved the sum completion time minimization problem for a $K$-user symmetric GMAC. Theorem \ref{theo:macweightedmini} can be thought as a generalization of their result, when $K=2$, to a general GMAC with asymmetric weights. Specializing Theorem \ref{theo:macweightedmini} with $P_1=P_2$ and $w=0.5$, we restate the result of \cite{Rai} in the following corollary.

\begin{corollary}
The solution of the weighted sum completion time minimization problem (\ref{eq:optoverall}) with equal weights for a symmetric GMAC, is given by
\begin{align*}
\bd^*=\begin{cases}
    \bd_D, \quad \tfrac{\tau_2}{\tau_1}\leq 1 \\
    \bd_E, \quad \tfrac{\tau_2}{\tau_1}\geq 1
\end{cases},
\end{align*}
where $\mathbf{d}_D=( \varphi_1, \ \tfrac{\tau_2}{\gamma(P_2)} )$ and $\mathbf{d}_E=( \tfrac{\tau_1}{\gamma(P_1)},\ \varphi_2 )$. Note that for $\frac{\tau_2}{\tau_1}=1$, $\mathbf{d}_D$ and $\mathbf{d}_E$ both minimize (\ref{eq:optoverall}).
\end{corollary}

For the problem of minimizing the weighted sum completion time in a two-user GMAC, it is clear that the optimal strategy depends on the amounts of data to be transmitted by each user, i.e. $\tau_i$, $i\in\{1,2\}$. It is also natural for one to expect that it is the ratio of $\tau_1$ and $\tau_2$, rather than the absolute values, that dictates the solution. Indeed, for the symmetric case, the optimal strategy is a communication analogue of the shorter-tasks-faster service policy \cite{Rai}, where the user with less data get a higher rate and is finished earlier. The intuition behind this simple strategy is that having a user finished earlier is not only beneficial to minimizing the delay for that user, but also preferable for the other user due to decreased interference in the remaining transmission time. It gets more involved in the asymmetric case, where not only does $\tfrac{\tau_2}{\tau_1}$ matter, but also the user powers and weights. Theorem \ref{theo:macweightedmini} gives the precise formula how these quantities interact with each other to determine the optimal strategy. It says that one should first decide which case the channel falls into using $\tfrac{\tau_2}{\tau_1}$ and the power. Then depending on the weights, either $\mathbf{d}_D$ or $\mathbf{d}_E$, one of the corners of the completion time region of a GMAC, minimizes the weighted sum completion time and the exact relation is given by Table \ref{table:macweightedmini}. Note that $\mathbf{d}_D$ and $\mathbf{d}_E$ are the completion times corresponding to the two corners of the GMAC capacity region $\mathbf{r}_D$ and $\mathbf{r}_E$ in Fig. \ref{fig:MACThreeCase}.

\subsection{Weighted Sum Completion Time Minimization in a GBC}
Since $\mathcal{D}^*$ of a GBC is not convex, it is more convenient to minimize the weighted sum completion time over the convex subregions $\mathcal{D}^*_i$ and then consider the overall minimum. Consider the following optimization problem
\begin{align}
    &\textrm{minimize}\quad\, d_{s,i}=wd_1+\bar{w}d_2 \label{eq:optorigin}\\
    &\textrm{subject to}\quad (d_1,d_2)\in\mathcal{D}_{i}^*,\ i=1,2 \notag
\end{align}
where $\mathcal{D}^*$ is the completion time region of the GBC in (\ref{channelGBC}) given in Corollary \ref{theo:CTRGBC} and $\mathcal{D}^*_i$ is defined in (\ref{def:ctrsubregion}). Notice that the expression of $\mathcal{D}^*$ is fairly complex, which makes a direct approach, i.e. finding the supporting line of $\mathcal{D}^*$, tedious. For this reason, we will use the fact that $G_i$ is an affine map (see Remark \ref{prop:linearity}) and tackle the problem in the rate domain. Before we state the main theorem, let us consider the following notations.

Let an arbitrary boundary point of the GBC capacity region be denoted by $\tilde{\mathbf{r}}(P_1)=(\tilde{r}_1(P_1), \tilde{r}_2(P_1))$,
\begin{align}
\label{GBCboundarypoint}
  \tilde{r}_1(P_1)=\gamma(h_1P_1),\quad \tilde{r}_2(P_1)=\gamma(h_2P)-\gamma(h_2P_1),
\end{align}
for some $P_1\in[0,P]$. Let us define
\begin{align}
\label{tangenteq}
    g(P_1)=\frac{1/h_1+P_1}{1/h_2+P_1}, \quad a_1(P_1)= \frac{g(P_1)}{\tilde{r}_2(P_1)+g(P_1)\tilde{r}_1(P_1)}, \quad a_2(P_1)=\frac{1}{\tilde{r}_2(P_1)+g(P_1)\tilde{r}_1(P_1)},
\end{align}
and two functions:
\begin{align}
\label{twofunc}
 \kappa_1(P_1)=1-a_2(P_1)r_2^*,\quad \kappa_2(P_1)= a_1(P_1)r_1^*,
\end{align}
where $r_i^*=\gamma(h_iP)$, $i\in\{1,2\}$, and $P_1\in[0,P]$.

\begin{lemma}
\label{lem:tangent}
For $\tilde{\mathbf{r}}(P_1)$ given by (\ref{GBCboundarypoint}), $a_i(P_1)$ given by (\ref{tangenteq}), and $\kappa_i(P_1)$ given by (\ref{twofunc}), $i\in\{1,2\}$, the following statements are true.
\begin{enumerate}
 \item If $a_1r_1+a_2r_2=1$ is the tangent line to the boundary of the GBC capacity region at $\tilde{\mathbf{r}}(P_1)$, then $a_1=a_1(P_1)$ and $a_2=a_2(P_1)$.
\item $\kappa_1(P_1)$ and $\kappa_2(P_1)$ in (\ref{twofunc}) are strictly increasing functions of $P_1\in[0,P]$. Furthermore $\kappa_1(P_1)\in[0,1)$ and $\kappa_2(P_1)\in(0, 1]$ for any $P_1\in[0,P]$.
\end{enumerate}
\end{lemma}
\begin{IEEEproof}
See Appendix \ref{proof:lem:tangent}.
\end{IEEEproof}

Referring to Fig. \ref{fig:BCCTR}(a), $\mathbf{r}_C$ denotes the intersection of line $\frac{r_2}{r_1}=\frac{\tau_2}{\tau_1}$ and the boundary of the GBC capacity region. For any given $\frac{\tau_2}{\tau_1}$, $\mathbf{r}_C=(r_{1,C},\frac{\tau_2}{\tau_1}r_{1,C})$, $r_{1,C}=\gamma(h_1P_{1,C})$, can be solved numerically for some $P_{1,C}\in[0,P]$, i.e $\tilde{\mathbf{r}}(P_{1,C})=\mathbf{r}_C$. Let $\mathbf{r}_A$ and $\mathbf{r}_B$ denote the $r_2$-axis and $r_1$-axis intercepts of the boundary of the GBC capacity region respectively. Then $\mathbf{r}_A=(0,\gamma(h_2P))$, i.e. $\tilde{\mathbf{r}}(0)=\mathbf{r}_A$ and $\mathbf{r}_B=(\gamma(h_1P),0)$, i.e. $\tilde{\mathbf{r}}(P)=\mathbf{r}_B$. Due to Lemma \ref{lem:tangent}, we have $0\leq \kappa_1(P_{1,C})\leq \kappa_1(P)<1$ and for an arbitrary weight $w\in(\kappa_1(P_{1,C}),\kappa_1(P))$, the solution of $\kappa_1(P_1)=w$ can be denoted by $P_1=\kappa_1^{-1}(w)$, where $\kappa_1^{-1}$ is the inverse of $\kappa_1$. Note that $\kappa_1^{-1}$ is well-defined because of the strict monotonicity of $\kappa_1$ shown Lemma \ref{lem:tangent}, however in general it cannot be expressed in closed form and can only be determined numerically. Similarly for an arbitrary weight $w\in(\kappa_2(0), \kappa_2(P_{1,C}))$, the unique solution of $\kappa_2(P_1)=w$ can be denoted by $P_1=\kappa_2^{-1}(w)$, where where $\kappa_2^{-1}$ is the inverse of $\kappa_2$.

\begin{theorem}
\label{theo:miniBC}
Let
\begin{align*}
  \kappa_1(P)&=1-\tfrac{(1/h_2+P)\gamma(h_2P)}{(1/h_1+P)\gamma(h_1P)},\quad \kappa_1(P_{1,C})=1- \tfrac{\tau_1(1/h_2+P_{1,C})\gamma(h_2P)}{[\tau_1(1/h_1+P_{1,C}) +\tau_2(1/h_2+P_{1,C})]\gamma(h_1P_{1,C})},\\
  \kappa_2(0)&=\tfrac{\gamma(h_1P)h_2}{\gamma(h_2P)h_1},\quad \kappa_2(P_{1,C})=\tfrac{\tau_1(1/h_1+P_{1,C})\gamma(h_1P)}{[\tau_1(1/h_1+P_{1,C}) +\tau_2(1/h_2+P_{1,C})] \gamma(h_1P_{1,C})},
\end{align*}
where $P_{1,C}\in[0,P]$ such that $\tilde{\mathbf{r}}(P_{1,C})=(\tilde{r}_{1} (P_{1,C}), \frac{\tau_2}{\tau_1} \tilde{r}_{1} (P_{1,C}))$. Denote $\mathbf{d}_A=G_2(\mathbf{r}_A)=(\frac{\tau_1}{\gamma(h_1P)}+\frac{\tau_2}{\gamma(h_2P)},\frac{\tau_2}{\gamma(h_2P)})$, $\mathbf{d}_B=G_1(\mathbf{r}_B)=(\frac{\tau_1}{\gamma(h_1P)},\frac{\tau_1}{\gamma(h_1P)}+\frac{\tau_2}{\gamma(h_2P)})$, $\mathbf{d}_C=G_i(\mathbf{r}_C)=G_i(\tilde{\mathbf{r}}(P_{1,C}))$, $\mathbf{d}_i(w)=G_i(\tilde{\mathbf{r}}(\kappa_i^{-1}(w)))$, $i\in\{1,2\}$, where $G_i$ is given by Lemma \ref{lem:mapping}, $\tilde{\mathbf{r}}$ is given by (\ref{GBCboundarypoint}), and $\kappa_i^{-1}$ is the inverse of $\kappa_i$ given by (\ref{twofunc}).

For the GBC in (\ref{channelGBC}), the solution of the weighted sum completion time minimization problem (\ref{eq:optoverall}) is given by $d_s^*=\min\{d_{s,1}^*,d_{s,2}^*\}$, where $d_{s,i}^*$ is the solution of the weighted sum completion time minimization problem  defined over subregion $\mathcal{D}_i^*$ (\ref{eq:optorigin}) and is achieved by the completion time pair given by Table \ref{table:bcweightedmini}.
\end{theorem}
\begin{IEEEproof}
See Appendix \ref{proof:theo:theo:miniBC}.
\end{IEEEproof}

\begin{center}
\begin{table}[htbp]
\caption{Solution to the weighted sum completion time minimization problem for a GBC}
\label{table:bcweightedmini}
\hfill{}
\begin{tabular}{ll}
\hline\hline
   $\qquad \qquad i=1$       &       $\qquad \qquad i=2$ \\ \hline
$\mathbf{d}_C, w\in[0,\kappa_1(P_{1,C})]$     &     $\mathbf{d}_A, w\in[0,\kappa_2(0)]$  \\
$\mathbf{d}_1(w) ,\, w\in(\kappa_1(P_{1,C}),\kappa_1(P))$    &     $ \mathbf{d}_2(w), \, w\in(\kappa_2(0), \kappa_2(P_{1,C}))$ \\
$\mathbf{d}_B, w\in[\kappa_1(P),1]$        &       $\mathbf{d}_C, w\in[\kappa_2(P_{1,C}),1]$ \\
\hline
\end{tabular}
\hfill{}
\end{table}
\end{center}
\begin{remark}
The presentation of the optimal solution to the weighted sum completion time minimization problem for a GBC is considerably more complicated than the case of a GMAC. This is partially because equations $\tilde{\mathbf{r}}(P_{1,C})=(\tilde{r}_{1}(P_{1,C}),\frac{\tau_2}{\tau_1}\tilde{r}_{1}(P_{1,C}))$ and $\kappa_i(P_1)=w$, $i\in\{1,2\}$, can only be solved numerically for $P_{1,C}$ and $P_1$ respectively. However we note that these computations are fairly easy. To interpret Theorem \ref{theo:miniBC}, let us consider $i=1$. For given $\frac{\tau_2}{\tau_1}$ and $P$, let us assume $P_{1,C}$ is determined, so is $\mathbf{r}_C$. If the priority of user 2 is high enough, i.e. $w$ is small enough say less than $\kappa_1(P_{1,C})$, then the optimal operating rate $\mathbf{r}^*$ in the first phase when both users are active is $\mathbf{r}_C$. In the other extreme, if the priority of user 1 is high enough, i.e. $w\geq \kappa_1(P)$, then $\mathbf{r}^*=\mathbf{r}_B$. For any $w\in(\kappa_1(P_{1,C}),\kappa_1(P))$, a unique point on the boundary of the capacity region, i.e. curve $\widetilde{\mathbf{r}_B\mathbf{r}_C}$, is optimal and this point is given by $\mathbf{r}^*=\tilde{\mathbf{r}}(\kappa_1^{-1}(w))$. After the completion of one user in the first phase, the remaining user transmits at the maximum point-to-point rate in the second phase. The overall completion time pair can be expressed as $\mathbf{d}^*=G_1(\mathbf{r}^*)$.
\end{remark}

We now use the above result to prove the non-convexity of the GBC completion time region.
\begin{proposition}
\label{prop:BCnoconvex}
The completion time region of the GBC in (\ref{channelGBC}) given by Corollary \ref{theo:CTRGBC} is not convex.
\end{proposition}
\begin{IEEEproof}
See Appendix \ref{proof:prop:BCnoconvex}.
\end{IEEEproof}

\section{Completion Time Region in Gaussian Channels with Expected Block Power Constraint}
In the completion time region computation carried out for Gaussian channels such as a GMAC, a GBC and a GIC, we considered an expected per symbol power constraint for the input signal. If instead we have an expected block power constraint, because of the multi-phase operation detailed in Lemma \ref{lem:acheivecrate} and the possibility of power allocation among the phases, the computation of the completion time region becomes more involved and unlike Section III and V, the completion time region cannot be simply found by mapping the capacity region. In order to illustrate this, in what follows we will mainly focus on a GMAC. A brief discussion on other Gaussian channels will be provided at the end of this section.

\begin{definition}
For the GMAC given in (\ref{channelGMAC}), the expected block power constraint is defined as
\begin{align}
\label{eq:cost}
    \frac{1}{n_i} \sum_{t=1}^{n_i}E(X_{i,t}^2) \leq P_i,\quad i\in\{1,2\}.
\end{align}
\end{definition}

Note that the cost is only counted for the actual number of channel uses where each user is active. Recall that the capacity region $\mathcal{C}$ of the GMAC in (\ref{channelGMAC}) is given by (\ref{eq:GMACCapa}), which is a function of transmitter powers $P_1$, $P_2$. In the following, to emphasize this point, we use $\mathcal{C}(P_1,P_2)$ to refer to $\mathcal{C}$.

\begin{corollary}
\label{coro:GMACcost}
The $c$-constrained capacity region $\mathfrak{C}_c$ of the GMAC in (\ref{channelGMAC}) with expected block power constraint (\ref{eq:cost}) is the set of all non-negative $(R_1,R_2)$ satisfying
\begin{enumerate}
\item for $c\leq 1$ and all $P_{2,1},P_{2,2}\geq 0$ such that $cP_{2,1}+(1-c)P_{2,2}=P_2$,
\begin{align*}
  (R_1,[\tfrac{1}{c}R_2-(\tfrac{1}{c}-1)\gamma(P_{2,2})]^+)\in\mathcal{C}(P_1,P_{2,1}),
\end{align*}
\item for $c\geq 1$ and all $P_{1,1},P_{1,2}\geq 0$ such that $\frac{1}{c}P_{1,1}+(1-\frac{1}{c})P_{1,2}=P_1$,
\begin{align*}
  ([cR_1-(c-1)\gamma(P_{1,2})]^+,R_2)\in\mathcal{C}(P_{1,1},P_2),
\end{align*}
\end{enumerate}
where $\mathcal{C}(P_1,P_{2,1})$ ($\mathcal{C}(P_{1,1},P_2)$) is given by (\ref{eq:GMACCapa}) with $P_2$ ($P_1$) replaced by $P_{2,1}$ ($P_{1,1}$).
\end{corollary}

\begin{IEEEproof}
The proof is similar to that of Theorem \ref{theo:network} and is omitted.
\end{IEEEproof}

Let us focus on $c\leq 1$. Similar to Lemma \ref{lem:mapping}, we can obtain the following map
\begin{align*}
    d_1=\frac{\tau_1}{r_1},\quad d_2=\frac{\tau_2}{\gamma(P_{2,2})}+\frac{\left[\gamma(P_{2,2})-r_2\right]\tau_1}{\gamma(P_{2,2})r_1},
\end{align*}
for some $(r_1,r_2)\in\mathcal{C}_1(P_1,P_{2,1})=\mathcal{C}(P_1,P_{2,1})\bigcap \{(r_1,r_2): \tfrac{r_2}{r_1}\leq \tfrac{\tau_2}{\tau_1} \}$ and some $P_{2,1}, P_{2,2}$ such that $cP_{2,1}+(1-c)P_{2,2}=P_2$. Comparing this with (\ref{lem:mapping:eqd1}), the expected block power constraint (\ref{eq:cost}) brings in a new dimension for optimization, that is how to optimally distribute the power budget among the $c$ and $1-c$ fractions of time, which in return depends on $d_1,d_2$ through $c=\frac{d_1}{d_2}$. This presents an issue similar to the one when trying to determine the completion time region $\mD^*$ by directly using the constrained capacity region, discussed in Section II-D. Hence $\mD^*$ cannot be simply found by mapping the capacity region. In general there does not exist a closed-form expression for $\mD^*$. In the following, efforts will be dedicated to determine $\mD^*$ numerically.

\begin{figure}[htb]
  \centering
  \includegraphics[width=55mm]{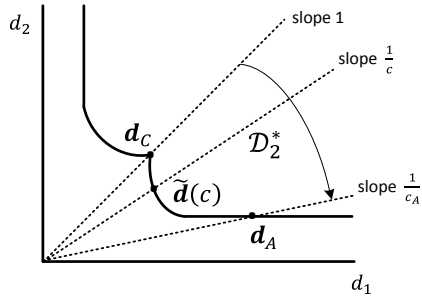}
  \caption{The completion time region of a GMAC with expected block power constraint}
  \label{fig:regionwithcost}
\end{figure}

Same as before, we compute subregion $\mD_i^*$, $i\in\{1,2\}$, individually and then take the union to obtain $\mathcal{D}^*$. In the following, we focus on $\mD_2^*$. In a similar fashion, $\mD_1^*$ can be obtained. First of all, we note that the structural property of $\mD^*$ given in Theorem \ref{theo:networkctr} carries over to the case with expected block power constraint. Specifically, referring to Fig. \ref{fig:regionwithcost}, the boundary of $\mathcal{D}_2^*$ consists of a 45-degree ray emanating from $\bd_C$, a horizontal ray emanating from $\bd_A$ and curve $\widetilde{\bd_A\bd_C}$. Next, we present a two-step procedure to compute the boundary of $\mathcal{D}_2^*$.

{\em Step 1: Determine the two rays}

Any point on the 45-degree ray arises as a result of the two users completing at the same time, i.e. $c=\frac{d_1}{d_2}=1$. Hence $\bd_C$ can be determined by mapping (using $G_2$) from the rate point where line $\frac{r_2}{r_1}=\frac{\tau_2}{\tau_1}$ intersects the boundary of $\mC$ in (\ref{eq:GMACCapa}). Thus the 45-degree ray emanating from $\bd_C$ is determined. Now let us consider $\bd_A$ given by: $d_{A,1}=d_{A,2}+\tfrac{\tau_1}{\gamma(P_1)}$, $d_{A,2}=\tfrac{\tau_2}{\gamma(P_2)}$, which is achieved by user 1 starting to transmit only after user 2 has finished. Note that the purpose of determining a point on the horizontal ray is to delimit the range of $c$, which will be swept numerically in step 2. Hence any point on the horizontal ray suffices and we choose $\bd_A$ as such due to its simplicity. Thus the horizontal ray emanating from $\bd_A$ is determined.

{\em Step 2: Determine curve $\widetilde{\bd_A\bd_C}$}

For each fixed $c\in[1,c_A]$, where $c_A=\frac{d_{A,1}}{d_{A,2}}$ and $d_{A,1}$, $d_{A,2}$ are found in step 1, referring to Fig. \ref{fig:regionwithcost} if we draw a line from the origin with slope $\frac{1}{c}$ and denote the intersection of this line and the boundary of $\mathcal{D}_2^*$ by $\tilde{\mathbf{d}}(c)$, then we can write $\tilde{\mathbf{d}}(c)=(\frac{\tau_1}{R_{1}^*(c)},\frac{\tau_2}{R_{2}^*(c)})$, where $\mathbf{R}^*(c)=(R_{1}^*(c),R_{2}^*(c))$ denotes the intersection of line $\frac{R_2}{R_1}=\frac{\tau_2}{\tau_1}c$ and the boundary of $\mathfrak{C}_c$ given in Corollary \ref{coro:GMACcost}. Next we show how to compute $\mathbf{R}^*(c)$ for each $c\in[1,c_A]$. We denote the power policy as $\mathcal{P}(c)=\{(P_{1,1},P_{1,2}): P_{1,1}\geq 0,\ P_{1,2}\geq 0,\ \frac{1}{c}P_{11}+(1-\frac{1}{c})P_{1,2}=P_1 \}$. We use $\mathfrak{R}_c(\mathcal{P}(c))$ to denote a $c$-constrained rate region that has the same expression as given in Corollary \ref{coro:GMACcost}, but for an arbitrary $\mathcal{P}(c)$, and denote the intersection of line $\frac{R_2}{R_1}= \frac{\tau_2}{\tau_1}c$ and  $\mathfrak{R}_c(\mathcal{P}(c))$ boundary by $\mathbf{R}(\mathcal{P}(c))$. Then $\mathbf{R}^*(c)=\mathbf{R}(\mathcal{P}^*(c))$, where $\mathcal{P}^*(c)$ is the optimal power policy that maximizes elements of $\mathbf{R}(\mathcal{P}(c))$ (Note that the two elements are maximized simultaneously by the same $\mathcal{P}^*(c)$). Because of the explicit expression of $\mathfrak{R}_c(\mathcal{P}(c))$, $\mathcal{P}^*(c)$ can be determined easily (e.g. numerical search through discretized power allocations). In this way we can determine, for a fixed $c\in[1,c_A]$, one point on curve $\widetilde{\bd_A\bd_C}$, i.e. $\tilde{\mathbf{d}}(c)$ where line $\frac{d_1}{d_2}=c$ crosses. Now let $c$ sweep through $[1,c_A]$, we can trace every point on $\widetilde{\bd_A\bd_C}$.

\begin{figure}[htb]
  \centering
  \psfrag{d1}[c][c][0.8][0]{$d_1$}
  \psfrag{d2}[c][c][0.8][0]{$d_2$}
  \psfrag{E}[r][l][0.4][0]{}
  \psfrag{D}[r][l][0.4][0]{}
  \includegraphics[width=55mm]{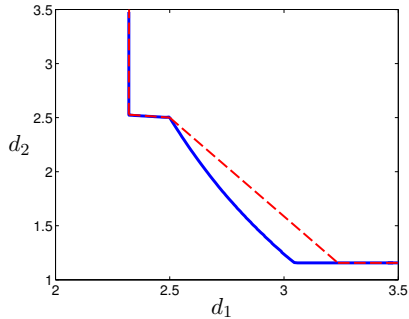}
  \caption{$\mathcal{D}^*$ with expected per symbol power constraint (dashed line) vs. $\mathcal{D}^*$ with expected block power constraint (solid line)}
  \label{fig:MACcost}
\end{figure}
An example is provided in Fig. \ref{fig:MACcost}, where $P_1=5$, $P_2=10$, $\tau_1=3$ and $\tau_2=2$. It is observed that $\mathcal{D}^*$ with expected block power constraint includes $\mD^*$ with expected per symbol power constraint given in Corollary \ref{theo:GMACCTR}. This is expected since the latter can be viewed as a special case of the former where $P_{i,1}=P_{i,2}=P_i$ for $i\in\{1,2\}$.

\begin{remark}
The completion time region with expected block power constraint can be pursued for other two-user Gaussian channels as well. As an example, let us consider a GBC with an expected block power constraint: $\frac{1}{n_{\pi_2}}\sum_{t=1}^{n_{\pi_2}}E(X_t^2) \leq P$. We can rewrite this as (assuming $c\leq 1$, i.e. $\pi_2=2$)
\begin{align*}
      \frac{1}{n_{2}}\sum_{t=1}^{n_{2}}E(X_t^2)
    = c \left[\frac{1}{n_{1}}\sum_{t=1}^{n_{1}}E(X_t^2) \right]+ (1-c)\left[\frac{1}{n_{2}-n_{1}}\sum_{t=n_{1}+1}^{n_{2}}E(X_t^2) \right] \leq P,
\end{align*}
which resembles the two-phase operation. We assign each phase with power $P_i$, $i\in\{1,2\}$, such that $cP_1+(1-c)P_2= P$. During each phase, the optimal strategy is to transmit at constant power. Hence following the same two-step procedure discussed above for a GMAC, we can compute $\mathcal{D}^*$ with expected block power constraint for a GBC.
\end{remark}

\section{Conclusions and Discussions}
\subsection{Summary}
In a channel where multiple users, each with a non-replenishable bit pool, compete channel resources to transmit their data in the shortest amount of time, users may benefit from decreased multi-user interference if others have already completed their data transmission. To capture this in an information-theoretic setting, in this paper we have studied the completion time problem for the two user case. The notion of completion time was formulated based on the concept of constrained rates, where codewords for different users need not be of the same length. Analogous to the capacity region, the completion time region characterizes all possible trade-offs between users' completion times and has been established for a DM-MAC. For Gaussian channels with expected per symbol power constraint, including a GMAC, a GBC and a GIC, completion time region or inner and outer bounds have also been obtained. When an utility optimization problem is defined over the completion time region and solved, the information-theoretic optimality of the resultant solution is assured. One example, minimizing the weighted sum completion time, has been solved for a GMAC and a GBC. For Gaussian channels with expected block power constraint, the completion time region cannot be found by simply mapping the capacity region and a numerical approach for the computation of $\mathcal{D}^*$ has been proposed.

\subsection{Completion Time in Multi-User Channels With Relays}
While this paper has dealt with the most common multi-user channels, i.e. MAC, BC and IC, one notable piece from the network information theory remains missing, i.e. relays. Below we give an example where the multi-phase operation, which leads to the characterization of the completion time region in all channels considered in this paper, ceases to be optimal with the presence of relays.

\begin{figure}[htb]
  \centering
  \includegraphics[width=55mm]{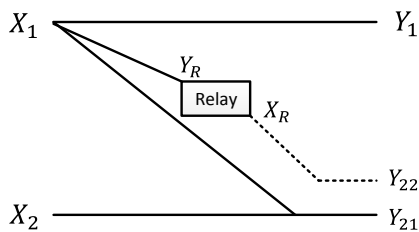}
  \caption{Gaussian Z interference channel with a relay}
  \label{fig:multihop}
\end{figure}

Consider the Gaussian interference-relay channel shown in Fig. \ref{fig:multihop}, where a half-duplex relay receives in-band and transmits out-of-band.
\begin{align}
  Y_1&=X_1+Z_1,\notag \\
  Y_R&=X_1+Z_R,\notag \\
  Y_{21}&=X_1+X_2+Z_{21},\quad Y_{22}=X_R+Z_{22}, \label{ZRIC}
\end{align}
where $Z_1,Z_R,Z_{21},Z_{22}$ are all i.i.d. Gaussian with variance 1 and $E(X_i^2)\leq P=1$, $i\in\{1,2\}$, $E(X_R^2)\leq P_R=0.25$. For $\tau_2=3\tau_1=3\tau$, we will show that the two-phase operation --- the multi-user transmission phase followed by a point-to-point transmission phase --- is suboptimal.

First of all, we argue that $\mathbf{d}=(2\tau,6\tau)$ is an achievable completion time pair. To achieve $d_1=2\tau$, source 1 transmits at rate $\gamma(P)=0.5$ until completion. To achieve $d_2=6\tau$, source 2 also needs to transmit at rate $\gamma(P)$ and this is possible if interference can be removed completely, which is accomplished as follows. The relay decodes interference message in the first $2\tau$ channel uses, re-encodes interference message and forwards for the remaining $4\tau$ channel uses. The total amount of mutual information accumulated at receiver 2 for interference decoding during the whole transmission is given by $2\tau\gamma(\frac{P}{1+P})+4\tau\gamma(P_R)$, where the first term is from treating receiver 2's own signal as noise in the first $2\tau$ channel uses and the second is from relay transmission during the remaining $4\tau$ channel uses. With $P=1$, $P_R=0.25$, this is greater than $2\tau\gamma(P)$, the amount of mutual information accumulated at receiver 1. Note that receiver 1 can only use the first $2\tau$ channel outputs to perform decoding in order to achieve $d_1=2\tau$, while receiver 2 can defer interference decoding until it receives all $4\tau$ channel outputs from relay transmission. These, combined with the first $2\tau$ channel outputs at receiver 2, jointly resolve interference message. Hence interference can be completely eliminated and $d_2=6\tau$ is achieved.

We next show that the sum capacity of channel (\ref{ZRIC}) is less than $2\gamma(P)$. It is easy to see $I(X_1;Y_1)\leq I(X_1,X_R;Y_{21},Y_{22}|X_2)$ for all $p(x_1)p(x_R)p(x_2)$. Similar to \cite{Costa}, this implies the multi-letter version $I(X_1^n;Y_1^n)\leq I(X_1^n,X_R^n;Y_{21}^n,Y_{22}^n|X_2^n)$. Then it is easy to show
\begin{align}
  n(R_1+R_2)-\epsilon_n&\leq I(X_1^n;Y_1^n)+I(X_2^n;Y_{21}^n,Y_{22}^n)\notag\\
&\leq nI(X_1,X_{R},X_{2};Y_{21},Y_{22})\notag\\
&\leq\gamma(2P+P_R+2PP_R) \label{neweq}\\
&<2\gamma(P),\label{neweq2}
\end{align}
where (\ref{neweq}) is due to the fact that Gaussian distribution maximizes entropy given covariance constraint and (\ref{neweq2}) is because of $P=1$ and $P_R=0.25$. Hence in an independent two-phase operation of the channel, it is impossible for both users to transmit at the maximum rate $\gamma(P)$ simultaneously in the first phase, where the decodings must take place by the end of the first phase. Consequently the constrained rates achieved in the two-phase operation is less than $(\gamma(P),\gamma(P))$, which is required to achieve $\mathbf{d}=(2\tau,6\tau)$. Therefore $\mathbf{d}=(2\tau,6\tau)$ is not achievable in the multi-phase scheme.

The above example illustrates that intermediate nodes introduce memory to the channel, which invalidates the optimality of multi-phase operation. Similarly transmitter/receiver cooperation or feedback, could also result in transmitted signals to be dependent on the messages of completed users, making it impossible to decouple different phases of transmissions without losing the optimality.

\appendices
\section{Proof of Lemma \ref{lem:tangent}}
\label{proof:lem:tangent}
To simplify the notation, here we drop the argument and simply use $\tilde{\mathbf{r}}$ to refer to $\tilde{\mathbf{r}}(P_1)$ given in the Lemma. $g$, $a_i$ and $\kappa_i$, $i\in\{1,2\}$, follow similarly. The slope of the tangent line at $\tilde{\mathbf{r}}$ is equal to $-\frac{a_1}{a_2}$ and we can show $\frac{a_1}{a_2}=g=\frac{1/h_1 + P_1 }{1/h_2 + P_1}$. Furthermore because $a_1\tilde{r}_1+a_2\tilde{r}_2=1$, we obtain $a_1= \frac{g}{\tilde{r}_2+g\tilde{r}_1}$ and $a_2=\frac{1}{\tilde{r}_2+g\tilde{r}_1}$. Hence item 1 is proved.

Writing $\frac{1}{a_2}=\tilde{r}_2+g\tilde{r}_1$, $\frac{1}{a_1}= \frac{\tilde{r}_2}{g}+\tilde{r}_1$, and using $\tilde{r}_1=\gamma(h_1P_1)$ and $\tilde{r}_2=\gamma(h_2P)-\gamma(h_2P_1)$, we can show (note $h_1\geq h_2$)
\begin{align*}
    \tfrac{d}{dP_1}\left(\tfrac{1}{a_1}\right)&=-\tfrac{(h_1-h_2)h_1h_2}{(h_2+h_1h_2P_1)^2} \left[\gamma(h_2P)-\gamma(h_2P_1)\right]<0 \text{ for }P_1\neq P\\
    \tfrac{d}{dP_1}\left(\tfrac{1}{a_2}\right)&=\tfrac{(h_1-h_2)h_1h_2}{(h_1+h_1h_2P_1)^2}\gamma(h_1P_1)>0 \text{ for }P_1\neq 0.
\end{align*}
Since $\frac{1}{a_1}$ is strictly decreasing and $\frac{1}{a_2}$ is strictly increasing, $\kappa_1$ and $\kappa_2$ are strictly increasing functions of $P_1\in[0,P]$. It is obvious $\kappa_1< 1$ and $\kappa_2> 0$, since $a_1,a_2>0$. Because $\frac{1}{a_1}$ is a strictly decreasing function of $P_1\in[0,P]$, we have the maximum $a_1=\frac{1}{r_1^*}$. Hence $\kappa_2\leq 1$. Similarly because $\frac{1}{a_2}$ is a strictly increasing function of $P_1\in[0,P]$, we have the maximum $a_2=\frac{1}{r_2^*}$. Hence $\kappa_1\geq 0$. Therefore item 2 is proved.

\section{Proof of Theorem \ref{theo:miniBC}}
\label{proof:theo:theo:miniBC}
We prove Theorem \ref{theo:miniBC} for $i=1$. The case of $i=2$ follows similarly. Suppose the solution of (\ref{eq:optorigin}) for $i=1$ is given by $d_{s,1}^*$. Hence line $wd_1+\bar{w}d_2=d_{s,1}^*$ supports $\mathcal{D}_1^*$. Let us first focus on the non-end-point case, i.e. the line is tangent to the boundary of $\mathcal{D}_1^*$ at some point other than $\mathbf{d}_B$ or $\mathbf{d}_C$ (see Fig. \ref{fig:BCCTR}). Because $G_1$ is an affine map from Remark \ref{prop:linearity}, the image of $wd_1+\bar{w}d_2=d_{s,1}^*$ under $G_1$ is a line in the capacity region and is given by $\frac{1}{\tau_1}\left[d_{s,1}^*-\frac{\bar{w}\tau_2}{r_2^*} \right]r_1 + \frac{\bar{w}}{r_2^*}r_2=1$. We claim that this line must be tangent to the boundary of $\mathcal{C}_1$, i.e. $\widetilde{\mathbf{r}_B\mathbf{r}_C}$. To see this, suppose it is secant, i.e. there are more than one point that are both on the line and in $\mathcal{C}_1$. This implies that there are more than one point that are both on line $wd_1+\bar{w}d_2=d_{s,1}^*$ and in $\mathcal{D}_1^*$, which contradicts with $wd_1+\bar{w}d_2=d_{s,1}^*$ being tangent. Since Lemma \ref{lem:tangent} gives the expression of the tangent line to the boundary of the capacity region, we must have $\frac{\bar{w}}{r_2^*}=a_2(P_1)$, i.e. $w=\kappa_1(P_1)$, which can be solved for $P_1=\kappa_1^{-1}(w)$. Hence the tangent point to the boundary of $\mathcal{C}_1$ is $\tilde{\mathbf{r}}(P_1)$ and the corresponding supporting point of $\mathcal{D}_1^*$ is $G_1(\tilde{\mathbf{r}}(P_1))$ with $P_1=\kappa_1^{-1}(w)$.

Now consider the end point case. Similarly, we can show that $\mathbf{d}_B$ is the minimizer of (\ref{eq:optorigin}) for $w=\kappa_1(P)$. For any $w>\kappa_1(P)$, $\mathbf{d}_B$ would still be the only point where line $wd_1+\bar{w}d_2=d_{s,1}^*$ supports $\mathcal{D}_1^*$. Therefore $\mathbf{d}_B$ is the minimizer of (\ref{eq:optorigin}) for $w\in[\kappa_1(P),1]$. Similarly, we can show that $\mathbf{d}_C$ is the minimizer of (\ref{eq:optorigin}) for $w\in[0,\kappa_1(P_{1,C})]$.

\section{Proof of Proposition \ref{prop:BCnoconvex}}
\label{proof:prop:BCnoconvex}
Consider minimizing the weighted sum completion time $wd_1+\bar{w}d_2$. From Theorem \ref{theo:miniBC}, when $w=\kappa_1(P_{1,C})$, the minimum value $d_{s,1}^*$ is achieved at point $\mathbf{d}_C$, at which line $wd_1+\bar{w}d_2=d_{s,1}^*$ is tangent to the boundary of the completion time region, i.e. curve $\widetilde{\mathbf{d}_B \mathbf{d}_C}$ in Fig. \ref{fig:BCCTR}(b). The slope of this tangent line is $s_1=\frac{w}{w-1}=\frac{\kappa_1(P_{1,C})} {\kappa_1(P_{1,C})-1}$. Similarly the slope of the tangent to $\widetilde{\mathbf{d}_A \mathbf{d}_C}$ at point $\mathbf{d}_C$ is $s_2=\frac{\kappa_2(P_{1,C})}{\kappa_2(P_{1,C})-1}$. Note that $s_1,s_2\leq 0$. From Lemma \ref{lem:tangent}, the tangent line to the boundary of the capacity region at point $\mathbf{r}_C$ in Fig. \ref{fig:BCCTR}(a) is given by $a_1(P_{1,C})r_1+a_2(P_{1,C})r_2 =1$. Since $(r_1^*,r_2^*)$, where $r_i^*=\gamma(h_iP)$ for $i\in\{1,2\}$, is outside the capacity region, we have $a_1(P_{1,C})r_1^*+a_2(P_{1,C})r_2^* >1$, i.e. $\kappa_1(P_{1,C})< \kappa_2(P_{1,C})$ and consequently $s_1>s_2$. From supporting hyperplane theorem \cite{Boyd}, if $\mathcal{D}^*$ is convex, then there exists a supporting line at every boundary point of $\mathcal{D}^*$, particularly $\mathbf{d}_C$. Suppose such line exists at $\mathbf{d}_C$, then its slope, denoted by $s$, must satisfy $s_1\leq s\leq 0$, so that it supports $\mathcal{D}_1^*$, and $s\leq s_2$, so that it supports $\mathcal{D}_2^*$. However this is not possible due to $s_1>s_2$. Therefore $\mathcal{D}^*$ is not convex.


\end{document}